\documentclass{article}

\usepackage{xcolor}
\definecolor{nosaka}{rgb}{0.0, 0.5, 0.0}
\definecolor{nosaka2}{rgb}{0.6, 0.0, 0.0}

\usepackage{url}

\usepackage{listings}
\usepackage{longtable}

\usepackage[utf8]{inputenc}

\usepackage{hyperref}
\usepackage{graphicx}
\usepackage{tikz}
\usetikzlibrary{decorations.pathmorphing}

\usepackage{amsmath}

\makeatletter
\newcommand{\subscripts}[3]{%
  \@mathmeasure\z@\displaystyle{#2}%
  \global\setbox\@ne\vbox to\ht\z@{}\dp\@ne\dp\z@
  \setbox\tw@\box\@ne
  \@mathmeasure4\displaystyle{\copy\tw@_{#1}}%
  \@mathmeasure6\displaystyle{{#2}_{#3}}%
  \dimen@-\wd6 \advance\dimen@\wd4 \advance\dimen@\wd\z@
  \hbox to\dimen@{}\mathop{\kern-\dimen@\box4\box6}%
}
\makeatother

\usepackage{mathrsfs}

\usepackage{tikz}
\usepackage{bm}
\usepackage{arydshln}
\usepackage{color}
\usepackage{ulem}
\usepackage{amssymb}

\usepackage{framed}
\usepackage{latexsym}

\usepackage{geometry}
\numberwithin{equation}{section}
\allowdisplaybreaks

\begin{document}

\newcommand{\hiduke}[1]{\hspace{\fill}{\small [{#1}]}}
\newcommand{\aff}[1]{${}^{#1}$}
\renewcommand{\thefootnote}{\fnsymbol{footnote}}

\begin{titlepage}
\begin{flushright}
{\footnotesize preprint SISSA 28/2020/FISI}
\end{flushright}
\begin{center}
{\Large\bf
$\text{SU}(N)$ $q$-Toda equations from mass deformed ABJM theory}\\
\bigskip\bigskip
\bigskip\bigskip
{\large Tomoki Nosaka\footnote{\tt nosaka@yukawa.kyoto-u.ac.jp}}\aff{1,2}\\
\bigskip\bigskip
\aff{1}: {\small
\it International School for Advanced Studies (SISSA), Via Bonomea 265, 34136 Trieste, Italy
}\\
\aff{2}: {\small
\it INFN Sezione di Trieste, Via Valerio 2, 34127 Trieste, Italy
}\\
\bigskip
\end{center}
\bigskip
\bigskip
\begin{abstract}
It is known that the partition functions of the $\text{U}(N)_k\times \text{U}(N+M)_{-k}$ ABJM theory satisfy a set of bilinear relations, which, written in the grand partition function, was recently found to be the $q$-Painlev\'e $\text{III}_3$ equation.
In this paper we have suggested that a similar bilinear relation holds for the ABJM theory with ${\cal N}=6$ preserving mass deformation for an arbitrary complex value of mass parameter, to which we have provided several non-trivial checks by using the exact values of the partition function for various $N,k,M$ and the mass parameter.
For particular choices of the mass parameters labeled by integers $\nu,a$ as $m_1=m_2=-\pi i(\nu-2a)/\nu$, the bilinear relation corresponds to the $q$-deformation of the affine $\text{SU}(\nu)$ Toda equation in $\tau$-form.
\end{abstract}

\bigskip\bigskip\bigskip

\end{titlepage}

\renewcommand{\thefootnote}{\arabic{footnote}}
\setcounter{footnote}{0}

\tableofcontents

\section{Introduction and Summary}
To understand the M-theory, a postulated eleven dimensional theory which unifies various ten dimensional superstring theories non-perturbatively through an extra one dimension, is an important theme of research in theoretical particle physics.
In the eleven dimensional supergravity there are BPS black brane solutions with dimension $2+1$ (M2-brane) and $5+1$ (M5-brane), each of which has the gravitational entropy scaling $\sim N^{3/2}$ and $\sim N^{3}$ \cite{Klebanov:1996un} with respect to their charges $N$.
These scalings of the entropy are in contrast to the scaling for the $3+1$ dimensional black brane in the ten dimensional type IIB supergravity (D3-brane) which can be naturally understood as the degrees of freedom of the D3-brane worldvolume theory originating from the open strings ending on the stack of $N$ D3-branes.
The construction of the worldvolume theory of the M2-branes/M5-branes which reproduce these scalings of the entropy, while preserving the required supersymmetry and being compatible with the string theory under the compactification of the M-theory direction, is an important step toward understanding the M-theory.

These theories have been mysterious for a long time.
About the M2-branes, however, there has been a series of progress using Chern-Simons matter theories starting with \cite{Schwarz:2004yj}, and finally the ${\cal N}=6$ $\text{U}(N)_k\times \text{U}(N)_{-k}$ superconformal Chern-Simons matter theory \cite{Hosomichi:2008jd,Aharony:2008ug} was proposed as the theory of $N$ M2-branes placed on the orbifold background $\mathbb{C}^4/\mathbb{Z}_k$.
Later the $S^3$ partition function of this theory was calculated in the M-theory limit $N\rightarrow\infty$ with $k$ kept finite \cite{Herzog:2010hf} by using the supersymmetric localization formula \cite{Kapustin:2009kz}, and it was found that the free energy $F=-\log Z_{S^3}$ behaves as $F\approx (\pi\sqrt{2k}/3)N^{3/2}$, which perfectly agrees with the result obtained from the supergravity \cite{Drukker:2010nc}.

The supersymmetric localization formula \cite{Kapustin:2009kz} allows us to rewrite the $S^3$ partition function of the ABJM theory into an $N$ dimensional matrix model, where the rewriting is exact even for finite $N$.
Hence this matrix model have been studied intensively \cite{Suyama:2009pd,Drukker:2009hy,Marino:2009jd,Drukker:2010nc,Drukker:2011zy,Suyama:2011yz,Fuji:2011km,Okuyama:2011su} to reveal the structure of the $1/N$ corrections which would be related to the quantum effects in the M-theory.
One of the largest breakthroughs in this direction was \cite{Marino:2011eh}, where it was found that the $S^3$ partition function of the ABJM theory has the same structure as the partition function of one dimensional quantum mechanical system of $N$ non-interacting fermions with the one-particle density matrix given as\footnote{
In \cite{Marino:2011eh} the one-particle density matrix for the ABJM theory is given in a Hermitian form ${\widehat\rho}=(2\cosh{\widehat x}/2)^{-1/2}(2\cosh{\widehat p}/2)^{-1}(2\cosh{\widehat x}/2)^{-1/2}$ instead of ${\widehat\rho}=(2\cosh{\widehat x}/2)^{-1}(2\cosh{\widehat p}/2)^{-1}$.
The two expressions are related by a similarity transformation, and hence they are equivalent to each other as long as we consider only the traces $\text{Tr}{\widehat\rho}^n$.
}
${\widehat\rho}=(2\cosh{\widehat x}/2)^{-1}(2\cosh{\widehat p}/2)^{-1}$.
Here ${\widehat x}$ and ${\widehat p}$ are the canonical position/momentum operators satisfying $[{\widehat x},{\widehat p}]=2\pi ik$.

The Fermi gas formalism and various computational techniques developed based on it \cite{Hatsuda:2012hm,Putrov:2012zi,Hatsuda:2012dt,Calvo:2012du,Hatsuda:2013gj} enabled us to study the $1/N$ perturbative corrections as well as the non-perturbative effects of order ${\cal O}(e^{-\sqrt{N/k}}, e^{-\sqrt{kN}})$ \cite{Drukker:2011zy} quantitatively at finite $k$.
Finally in \cite{Hatsuda:2013oxa} the large $N$ expansion of the $S^3$ partition function of the ABJM theory was completely determined including all of the non-perturbative effects.
Defining the grand potential $J(\mu)$ as $\sum_{N=0}^\infty e^{\mu N}Z(N)=:\sum_{n\in\mathbb{Z}} e^{J(\mu+2\pi in)}$, they found that the large $\mu$ non-perturbative effects of order ${\cal O}(e^{-\mu/k},e^{-\mu})$ in $J(\mu)$, which correspond to ${\cal O}(e^{-\sqrt{N/k}},e^{-\sqrt{kN}})$ corrections in the partition function, are precisely given by the topological string free energy and the Nekrasov-Shatashvili limit of the refined topological string free energy both on local $\mathbb{P}^1\times \mathbb{P}^1$.
Later this result was also generalized to the case with non-equal ranks $\text{U}(N)_k\times \text{U}(N+M)_{-k}$ \cite{Honda:2014npa}.

On the other hand, it was recently discovered that the partition function of the ABJM theory is also related to the Painlev\'e systems \cite{Bonelli:2016idi,Bonelli:2017ptp,Bonelli:2017gdk}.
They found \cite{Bonelli:2016idi} that the grand partition function of the ABJM theory coincides, under a particualr limit, to the Fredholm determinant type solution of the $\tau$ function of the Painlev\'e $\text{III}_3$ equation which was found in \cite{Zamolodchikov:1994uw}.
They further found \cite{Bonelli:2017ptp} that the exact relations between the partition functions of the rank deformed ABJM theory discovered in \cite{Grassi:2014uua} can be reorganized into the $\tau$-form of the $q$-deformed Painlev\'e $\text{III}_3$ equation.

Our motivation is to find other examples of the theories of M2-branes which enjoy the similar relation.
On one hand, such new examples may provide us a new way to characterize a class of $q$-difference equations including $q$-Painlev\'e systems from the viewpoint of the M2-branes on various background geometries.
Such generalization would be also useful in the other direction: solving the M-theory by using $q$-difference equations.
As a generalization of the ABJM theory, we can consider its mass deformation \cite{Hosomichi:2008jb,Gomis:2008vc} where the bifundamental chiral multiplets have supersymmetric mass term, whose partition function were also studied in the large $N$ limit \cite{Jafferis:2011zi,Anderson:2015ioa,Nosaka:2015bhf,Nosaka:2016vqf} together with the $1/N$ corrections \cite{Nosaka:2015iiw,Honda:2018pqa}.
Recently the partition function of this theory was found to be useful to understand the correlation functions of the ABJM theory even for the supersymmetry non-protected sectors by using the bootstrap technique \cite{Poland:2010wg,ElShowk:2012hu,El-Showk:2014dwa, Chester:2014fya}, which were hoped to give a new information about the M-theory \cite{Agmon:2017xes,Chester:2018aca,Binder:2018yvd,Binder:2019mpb,Chester:2020jay,Binder:2020ckj}.
The exact values of the partition functions for general $N$ or an exact relation among them which holds beyond the $1/N$ perturbation may provide further hints toward understanding the M-theory.

Indeed as a class of the generalization of the relation between the ABJM matrix model and $q$-$\text{PIII}_3$, it was already proposed that the $\tau$-function of the affine $\text{SU}(\nu)$ $q$-Toda equation also allows a Fredholm determinant type solution, together with the concrete expression for the matrix models \cite{Bonelli:2017ptp}, although it was not clear whether those matrix models can arise in the context of the worldvolume theories of the M2-branes.
In this paper we have found that this matrix model with a special choice of the parameters, can be realized as the $S^3$ partition function of the $\text{U}(N)_k\times \text{U}(N+M)_{-k}$ ABJM theory with the ${\cal N}=6$ preserving mass deformation \cite{Hosomichi:2008jd,Hosomichi:2008jb,Gomis:2008vc}, where the mass parameter is pure imaginary and identified with the rank $\nu$ as $m_1=m_2=-\pi i(\nu-2a)/\nu$ (see \eqref{ZkMNm1m2} for the normalization).
Here the rank difference $M$ is related to the time variable of the Toda system.
To this 3d partition function, various techniques which were used to solve the ABJM theory are applicable straighforwardly.
In particular, we find that the partition function can be written in so called ``open string formalism'' \cite{Matsumoto:2013nya}, which allows us to calculate the partition function exactly at finite $k,m_1,m_2,N,M$.
These exact values provide a non-trivial check that the matrix model proposed in \cite{Bonelli:2017ptp} is actually a solution of the $\text{SU}(\nu)$ $q$-Toda equation.

This paper is organized as follows.
After fixing our notation for the partition function of the mass deformed ABJM theory in section \ref{sec_mABJM}, in section \ref{sec_conjecture} we display a conjectural bilinear relation among the partition functions for $m_1=m_2$ \eqref{conjecturedmABJMbilinearidentity}.
We also compare our proposal with the known relation between the partition function of the ABJM theory without mass deformation and the $q$-Painlev\'e $\text{III}_3$ (or equivalently the affine $\text{SU}(2)$ $q$-Toda) system \cite{Bonelli:2017gdk} and its higher rank generalization \cite{Bonelli:2017ptp,Bershtein:2018srt}.
In section \ref{sec_openstringformalism}, we rewrite the partition function of the mass deformed ABJM theory in the open string formalism and calculate the exact values of the partition function.
By using these exact values we have checked that our proposal \eqref{conjecturedmABJMbilinearidentity} holds for a generic complex value of $m_1=m_2$ up to ${\cal O}(z^5)$ or higher for some special values of $k$.
In section \ref{sec_discuss} we summarize our results and discuss possible future works.
In appendix \ref{sec_proofCVdM} we review the proof of the Cauchy-Vandermonde determinant formula \eqref{CauchyVdM}.
In appendix \ref{app_listofexactvalues} we display the explicit expressions for some of the exact values of the partition function obtained by using the methods in section \ref{sec_openstringformalism}.

\section{Partition function of mass deformed ABJM theory}
\label{sec_mABJM}

The ABJM theory is an ${\cal N}=6$ superconformal Chern-Simons matter theory which consists of the two vector multiplets of the gauge groups $\text{U}(N)$,$\text{U}(N+M)$ with the Chern-Simons actions with the Chern-Simons levels $\pm k$, and the two pairs of the chiral multiplets in the bifundamental representation and the anti-bifundamental representation under the gauge group $\text{U}(N)\times \text{U}(N+M)$ \cite{Hosomichi:2008jd,Aharony:2008ug,Aharony:2008gk}.
We can further add the mass term to the these chiral multiplets while preserving ${\cal N}=2$ supersymmetry by turning on the vevs of the background vector multiplets of the flavor symmetries \cite{Freedman:2013oja}.
The partition function of the mass deformed ABJM theory can be calculated by the supersymmetric localization formula \cite{Kapustin:2009kz} as\footnote{
The results for $M=0$ in this paper overlap with those displayed in \cite{Nosaka:2015iiw} where we have defined the partition function with the same normalization as \eqref{ZkMNm1m2} under the following parameter identifications:
\begin{align}
k_1^{\text{\cite{Nosaka:2015iiw}}}=k,\quad\quad k_2^{\text{\cite{Nosaka:2015iiw}}}=-k,\quad \zeta_1^{\text{\cite{Nosaka:2015iiw}}}=-\frac{im_1}{\pi},\quad \zeta_2^{\text{\cite{Nosaka:2015iiw}}}=-\frac{im_2}{\pi}.
\end{align}
}
\begin{align}
Z_{k,M}(N,m_1,m_2)&=\frac{
(-1)^{MN+\frac{M(M-1)}{2}}
e^{\frac{NM(m_1+m_2)}{2}}}{N!(N+M)!}\int\frac{d^N\lambda}{(2\pi)^N}\frac{d^{N+M}{\widetilde\lambda}}{(2\pi)^{N+M}}
e^{\frac{ik}{4\pi}(\sum_{i=1}^N\lambda_i^2-\sum_{m=1}^{N+M}{\widetilde\lambda}_m^2)}\nonumber \\
&\quad \times \frac{
\prod_{i<j}^N(2\sinh\frac{\lambda_i-\lambda_j}{2})^2
\prod_{m<n}^{N+M}(2\sinh\frac{{\widetilde\lambda}_m-{\widetilde\lambda}_n}{2})^2
}{
\prod_{i=1}^N\prod_{j=1}^{N+M}
2\cosh\frac{\lambda_i-{\widetilde\lambda}_n-m_1}{2}
2\cosh\frac{{\widetilde\lambda}_n-\lambda_i-m_2}{2}
},
\label{ZkMNm1m2}
\end{align}
where $m_1,m_2$ are the vevs of the scalar components of the background vector multiplets \cite{Freedman:2013oja,Honda:2018pqa}, and $\lambda_i$,${\widetilde\lambda}_m$ are the vevs of the scalar components of the vectormultiplets which label the localization locus.
Here we have introduced the overall factor $e^{NM(m_1+m_2)/2}$ for later convenience.
For $M=0$ it is known that the partition function can be rewritten as the partition function of $N$ particle ideal Fermi gas \cite{Nosaka:2015iiw,Marino:2011eh}
\begin{align}
Z_{k,0}(N,m_1,m_2)=\frac{1}{N!}\int \frac{d^Nx}{(2\pi)^N}\det_{i,j}\langle x_i|{\widehat\rho}_0|x_j\rangle,
\label{FermigasatM0}
\end{align}
with the one particle density matrix ${\widehat\rho}$ given as\footnote{
Here we adopt the following notation and normalization:
\begin{itemize}
\item ${\widehat x},{\widehat p}$: canonical position/momentum operators with $[{\widehat x},{\widehat p}]=2\pi ik$,
\item $|x\rangle$: position eigenstate, $|p\rangle\!\rangle$: momentum eigenstate,
\item $\langle x|x'\rangle=2\pi \delta(x-x')$, $\langle p|p'\rangle=2\pi \delta(p-p')$, and correspondingly, $\int \frac{dx}{2\pi}|x\rangle\langle x|=\int\frac{dp}{2\pi}|p\rangle\!\rangle\langle\!\langle p|=1$,
\item $\langle x|p\rangle\!\rangle=\frac{1}{\sqrt{k}}e^{\frac{ixp}{2\pi k}}$, $\langle\!\langle p|x\rangle=\frac{1}{\sqrt{k}}e^{-\frac{ixp}{2\pi k}}$.
\end{itemize}
We also use the following formulas several times
\begin{align}
e^{-\frac{i}{4\pi k}{\widehat p}^2}e^{-\frac{i}{4\pi k}{\widehat x}^2}
f({\widehat p})
e^{\frac{i}{4\pi k}{\widehat x}^2}
e^{\frac{i}{4\pi k}{\widehat p}^2}
=f({\widehat x}),\quad
e^{-\frac{i}{4\pi k}{\widehat p}^2}
e^{-\frac{i}{4\pi k}{\widehat x}^2}
|p\rangle\!\rangle
=\frac{1}{\sqrt{i}}e^{\frac{i}{4\pi k}p^2}|p\rangle,
\label{|>>to|>henkan}
\end{align}
where in the second formula $|p\rangle\!\rangle$ is a momentum eigenstate, while $|p\rangle$ is a position eigenstate with the eigenvalue $p$.
\label{footnote1}
}
\begin{align}
{\widehat\rho}_0=
\frac{e^{-\frac{im_2}{2\pi}{\widehat x}}}{2\cosh\frac{{\widehat x}}{2}}
\frac{e^{-\frac{im_1}{2\pi}{\widehat p}}}{2\cosh\frac{{\widehat p}}{2}}.
\label{rho0}
\end{align}

We also define the grand partition function $\Xi_{k,M}(z,m_1,m_2)$ of the mass deformed ABJM theory as
\begin{align}
\Xi_{k,M}(z,m_1,m_2)=\sum_{N=0}^\infty z^N Z_{k,M}(N,m_1,m_2).
\label{XikM}
\end{align}
For $M=0$ the Fermi gas formalism \eqref{FermigasatM0} implies that $\Xi_{k,0}(z,m_1,m_2)$ is given as a Fredholm determinant
\begin{align}
\Xi_{k,0}(z,m_1,m_2)=\det(1+z{\widehat\rho}_0).
\end{align}

\section{Conjecture of bilinear identity for $Z_{k,M}(N,m_1,m_2)$}
\label{sec_conjecture}
In this paper we conjecture the following bilinear relation among the partition functions of the mass deformed ABJM theory with $m_1=m_2$:
\begin{align}
&\Xi_{k,M+1}(-e^{-m_1}z;m_1,m_1)
\Xi_{k,M-1}(-e^{m_1}z;m_1,m_1)
(1+e^{-\pi i(1-\frac{2M}{k})})\nonumber \\
&=\frac{Z_{k,M+1}(0)Z_{k,M-1}(0)}{Z_{k,M}(0)^2}
\Bigl[\Xi_{k,M}(z;m_1,m_1)^2
+e^{-\pi i(1-\frac{2M}{k})}
\Xi_{k,M}(-e^{-m_1}z;m_1,m_1)
\Xi_{k,M}(-e^{m_1}z;m_1,m_1)
\Bigr],
\label{conjecturedmABJMbilinearidentity}
\end{align}
where $\Xi_{k,M}(z;m_1,m_2)$ is the grand partition function defined as \eqref{XikM} and $Z_{k,M}(0)$ is the partition function of ${\cal N}=2$ $\text{U}(M)_{-k}$ pure Chern-Simons theory
\begin{align}
Z_{k,M}(0)=i^{\frac{M^2}{2}-M}e^{-\frac{\pi i M(M^2-1)}{6k}}k^{-\frac{M}{2}}\prod_{r>s}2\sin\frac{\pi(r-s)}{k}.
\label{ZkMpureCS}
\end{align}
This relation \eqref{conjecturedmABJMbilinearidentity} can be checked order by order in $z$ once we know the exact values of the partition functions $Z_{k,M}(N,m_1,m_2)$, which we compute in section \ref{sec_openstringformalism}.

Note that when $m_1=m_2=-\pi i(\nu-2a)/\nu$ with $\nu,a\in\mathbb{N}$ ($1\le a\le \nu-1$, $\nu,a$: coprime), \eqref{conjecturedmABJMbilinearidentity} is a higher rank generalization of the relation between the grand partition function of the ABJM theory and the affine $\text{SU}(2)$ $q$-Toda equation obtained in \cite{Bonelli:2017gdk}.
Indeed as we see below the partition function $Z_{k,M}(N,m_1,m_2)$ for $m_1=m_2$ coincides with the fermionic spectral trace of the inverse quantum mirror curve of $Y^{\nu,0}$ geometry \cite{Bonelli:2017ptp} whose relation to the affine $\text{SU}(\nu)$ $q$-Toda system was also argued in \cite{Bershtein:2018srt}.

\subsection{Closed string formalism and relation to $q$-Toda system}
\label{sec_closedstringformalism}
Our goal here is to rewrite the partition function $Z_{k,M}(N,m_1,m_2)$ \eqref{ZkMNm1m2} into the following form
\begin{align}
Z_{k,M}(N,m_1,m_2)=Z_{k,M}(0)\frac{1}{N!}\int \frac{d^N\lambda}{(2\pi)^N}\det_{i,j}\langle\lambda_i|{\widehat\rho}_M|\lambda_j\rangle,
\end{align}
with some one dimensional quantum mechanical operator ${\widehat\rho}_M$.\footnote{
This rewriting is called the closed string formalism \cite{Matsumoto:2013nya}, and was studied for the ABJM theory without mass deformation in \cite{Awata:2012jb,Honda:2013pea}.
}
Here $Z_{k,M}(0)$ is the partition function \eqref{ZkMNm1m2} for $N=0$, which is simply given as \eqref{ZkMpureCS}.
For this purpose, first of all we rewrite the one-loop determinant factors in the integrand of the partition function \eqref{ZkMNm1m2} by using the following Cauchy-Vandermonde determinant formula \cite{Matsumoto:2013nya} (see appendix \ref{sec_proofCVdM} for the proof)

\begin{align}
\frac{\prod_{i<j}^N2\sinh\frac{x_i-x_j}{2}\prod_{m<n}^{N+M}2\sinh\frac{y_m-y_n}{2}}{\prod_{i,m}2\cosh\frac{x_i-y_m}{2}}&=(-1)^{MN}e^{\frac{M}{2}\sum_{i=1}^Nx_i}e^{-\frac{M}{2}\sum_{n=1}^{N+M}y_n}
\det\begin{pmatrix}
\bigl[\frac{1}{2\cosh\frac{x_i-y_n}{2}}\bigr]_{i,n: N\times (N+M)}\\
[e^{\ell_r y_n}]_{r,n: M\times (N+M)}
\end{pmatrix},\nonumber \\
\frac{
\prod_{m<n}^{N+M}2\sinh\frac{y_m-y_n}{2}
\prod_{i<j}^N2\sinh\frac{x_i-x_j}{2}
}{
\prod_{m,i}2\cosh\frac{x_m-y_i}{2}
}
&=(-1)^{\frac{M(M-1)}{2}}e^{\frac{M}{2}\sum_{m=1}^{N+M}y_m}
e^{-\frac{M}{2}\sum_{j=1}^Nx_j}\nonumber \\
&\quad\det\Bigl(
\Bigl[\frac{1}{2\cosh\frac{y_m-x_j}{2}}\Bigr]_{m,j: (N+M)\times N}\,\,
[e^{-\ell_sy_m}]_{m,s: (N+M)\times M}
\Bigr),\label{CauchyVdM}
\end{align}
where $\ell_r=M+1/2-r$ with $r=1,2,\cdots,M$ as
\begin{align}
Z_{k,M}(N,m_1,m_2)&=\frac{1}{N!(N+M)!}\int\frac{d^N\lambda}{(2\pi)^N}\frac{d^{N+M}{\widetilde\lambda}}{(2\pi)^{N+M}}
\det\begin{pmatrix}
\bigl[\frac{1}{2k\cosh\frac{\lambda_i-{\widetilde\lambda}_n-m_1k}{2k}}\bigr]_{i,n}\\
\bigl[\frac{1}{\sqrt{k}}e^{\frac{\ell_r{\widetilde \lambda}_n}{k}}\bigr]_{r,n}
\end{pmatrix}\nonumber \\
&\quad
\det\Bigl(
\Bigl[e^{-\frac{i}{4\pi k}{\widetilde\lambda}_m^2}
\frac{1}{2k\cosh\frac{{\widetilde\lambda}_m-\lambda_j-m_2k}{2}}
e^{\frac{i}{4\pi k}\lambda_j^2}\Bigr]_{m,j}
\,\,
\Bigl[e^{-\frac{i}{4\pi k}{\widetilde\lambda}_m^2}\frac{1}{\sqrt{k}}e^{-\frac{\ell_s{\widetilde\lambda}_m}{k}}\Bigr]_{m,s}
\Bigr).
\end{align}
Here we have also rescaled the integration variables $\lambda_i,{\widetilde\lambda}_m$ as $(\lambda_i,{\widetilde\lambda}_m)\rightarrow (\lambda_i/k,{\widetilde\lambda}_m/k)$ for later convenience.
If we introduce the 1d quantum mechanical bra-ket notations (see footnote \ref{footnote1}), we can further rewrite the partition function as
\begin{align}
Z_{k,M}(N,m_1,m_2)&=\frac{1}{N!(N+M)!}\int\frac{d^N\lambda}{(2\pi)^N}\frac{d^{N+M}{\widetilde\lambda}}{(2\pi)^{N+M}}
\det\begin{pmatrix}
\bigl[\langle \lambda_i|\frac{e^{-\frac{im_1{\widehat p}}{2\pi}}}{2\cosh\frac{\widehat p}{2}}|{\widetilde\lambda}_n\rangle\bigr]_{i,n}\\
[\langle\!\langle 2\pi i\ell_r|{\widetilde\lambda}_n\rangle]_{r,n}
\end{pmatrix}\nonumber \\
&\quad \det\Bigl(
\Bigl[\langle{\widetilde\lambda}_m|e^{-\frac{i}{4\pi k}{\widehat x}^2}\frac{e^{-\frac{im_2{\widehat p}}{2\pi}}}{2\cosh\frac{\widehat p}{2}}
e^{\frac{i}{4\pi k}{\widehat x}^2}
|\lambda_j\rangle\Bigr]_{m,j}\,\,
[\langle{\widetilde\lambda}_m|e^{-\frac{i}{4\pi k}{\widehat x}^2}|2\pi i\ell_s\rangle\!\rangle
]_{m,s}
\Bigr).
\label{closed_first2}
\end{align}
Second we simplify the matrix elements of the second determinant by replacing
\begin{align}
|\lambda_i\rangle\rightarrow e^{\frac{i}{4\pi k}{\widehat p}^2}|\lambda_i\rangle,\quad
|{\widetilde \lambda}_m\rangle\rightarrow e^{\frac{i}{4\pi k}{\widehat p}^2}|{\widetilde \lambda}_m\rangle,
\end{align}
which results in
\begin{align}
\langle\lambda_i|\frac{e^{-\frac{im_1}{2\pi}{\widehat p}^2}}{2\cosh\frac{\widehat p}{2}}|{\widetilde\lambda}_n\rangle&\rightarrow
\langle\lambda_i|\frac{e^{-\frac{im_1}{2\pi}{\widehat p}^2}}{2\cosh\frac{\widehat p}{2}}|{\widetilde\lambda}_n\rangle,\nonumber \\
\langle\!\langle 2\pi i\ell_r|{\widetilde\lambda}_n\rangle&\rightarrow
e^{-\frac{\pi i\ell_r^2}{k}}\langle\!\langle 2\pi i\ell_r|{\widetilde\lambda}_n\rangle,\nonumber \\
\langle{\widetilde\lambda}_m|e^{-\frac{i}{4\pi k}{\widehat x}^2}\frac{e^{-\frac{im_2{\widehat p}}{2\pi}}}{2\cosh\frac{\widehat p}{2}}
e^{\frac{i}{4\pi k}{\widehat x}^2}
|\lambda_j\rangle&\rightarrow
\frac{e^{-\frac{im_2}{2\pi}\lambda_j}}{2\cosh\frac{\lambda_j}{2}} 2\pi \delta({\widetilde\lambda}_m-\lambda_j),
\nonumber \\
\langle{\widetilde\lambda}_m|e^{-\frac{i}{4\pi k}{\widehat x}^2}|2\pi i\ell_s\rangle\!\rangle&\rightarrow
\frac{1}{\sqrt{i}}e^{-\frac{\pi i\ell_s^2}{k}}2\pi \delta({\widetilde\lambda}_m-2\pi i\ell_s).
\end{align}
Here to obtain the last two expressions we have used the formulas \eqref{|>>to|>henkan}.
Third, after these replacements we further ``trivialize'' the second determinant in \eqref{closed_first2} as follows
\begin{align}
\int d^Nx\det_{i,j}[f_i(x_j)]\det_{i,j}[g_i(x_j)]
=N!\int d^Nx
\det_{i,j}[f_i(x_j)]
\Bigl(\prod_ig_i(x_i)\Bigr),
\label{trivialization}
\end{align}
to obtain
\begin{align}
Z_{k,M}(N,m_1,m_2)&=\frac{1}{N!}\int\frac{d^N\lambda}{(2\pi)^N}\frac{d^{N+M}{\widetilde\lambda}}{(2\pi)^{N+M}}
\Bigl(\prod_{r=1}^M e^{-\frac{\pi i\ell_r^2}{k}}\Bigr)
\det\begin{pmatrix}
\bigl[\langle \lambda _i|\frac{e^{-\frac{im_1}{2\pi}{\widehat p}}}{2\cosh\frac{\widehat p}{2}}|{\widetilde\lambda}_n\rangle\bigr]_{i,n}\\
\bigl[\langle\!\langle 2\pi i\ell_r|{\widetilde\lambda}_n\rangle\bigr]_{r,n}
\end{pmatrix}\nonumber \\
&\quad \Bigl(\prod_{i=1}^N\frac{e^{-\frac{im_2}{2\pi}\lambda_i}}{2\cosh\frac{\lambda_i}{2}}2\pi \delta({\widetilde\lambda}_i-\lambda_i)\Bigr)
\Bigl(\prod_{r=1}^M\frac{1}{\sqrt{i}}e^{-\frac{\pi i\ell_r^2}{k}}2\pi \delta({\widetilde\lambda}_{N+r}-2\pi i\ell_r)\Bigr).
\label{closed_third}
\end{align}
Now we can perform the ${\widetilde\lambda}_m$-integrations trivially just by replacing them according to the $\delta$-functions.
If we use the Cauchy-Vandermonde determinant formula \eqref{CauchyVdM} inversely we can rewrite the determinant factor in the first line of \eqref{closed_third} back into the product of hyperbolic functions.
This product can be rearranged into those which we can combine again into the Cauchy determinant \eqref{CauchyVdM} with $M=0$ and the rest.
Thus we finally obtain the following
\begin{align}
Z_{k,M}(N,m_1,m_2)&=Z_{k,M}(0)\frac{1}{N!}\int\frac{d^N\lambda}{(2\pi)^N}\det_{i,j}
\langle\lambda_i|
{\widehat \rho}_M
|\lambda_j\rangle,
\label{closedstringformalism}
\end{align}
with $Z_{k,M}(0)$ given as \eqref{ZkMpureCS} and 
\begin{align}
{\widehat \rho}_M&=(-1)^Me^{\frac{Mm_1}{2}}\frac{e^{-\frac{im_2}{2\pi}{\widehat x}}}{2\cosh\frac{{\widehat x}}{2}}
\prod_{r=1}^M\frac{2\sinh\frac{{\widehat x}-2\pi i\ell_r}{2k}}{2\cosh\frac{{\widehat x}-2\pi i\ell_r-m_1k}{2k}}
\frac{e^{-\frac{im_1}{2\pi}{\widehat p}}}{2\cosh\frac{{\widehat p}}{2}}.
\label{rhoMmABJM}
\end{align}

By using the quantum dilogarithm function $\Phi_b(x)$ \cite{Kashaev:2015wia}
\begin{align}
\Phi_b(x)&=\frac{\prod_{n=0}^\infty(1-e^{2\pi inb^2}e^{2\pi b(x+c_b)})}{\prod_{n=0}^\infty (1-e^{-2\pi inb^{-2}}e^{2\pi b^{-1}(x-c_b)})},\quad
c_b=\frac{i}{2}\Bigl(b+\frac{1}{b}\Bigr),
\end{align}
which satisfy the following recursive relation
\begin{align}
\frac{\Phi_b(x+inb)}{\Phi_b(x)}&=\frac{1}{\prod_{m=1}^n(1+e^{2\pi bx+2\pi ib^2(m-\frac{1}{2})})},\nonumber \\
\frac{\Phi_b(x+inb^{-1})}{\Phi_b(x)}&=\frac{1}{\prod_{m=1}^n(1+e^{2\pi b^{-1}x+2\pi ib^{-2}(m-\frac{1}{2})})},
\end{align}
the partition function \eqref{closedstringformalism} can be rewritten as ($b=\sqrt{k}$)
\begin{align}
Z_{k,M}(N,m_1,m_2)&=Z_{k,M}(0)\frac{1}{N!}\int \frac{d^N\lambda}{(2\pi)^N}\prod_ie^{(\frac{1}{2}-\frac{im_2}{2\pi})\lambda_i}
\frac{
\Phi_b(\frac{\lambda_i}{2\pi b}-\frac{iM}{b}+\frac{ib}{2})
\Phi_b(\frac{\lambda_i}{2\pi b}-\frac{m_1b}{2\pi})
}{
\Phi_b(\frac{\lambda_i}{2\pi b}-\frac{i}{b})
\Phi_b(\frac{\lambda_i}{2\pi b}-\frac{iM}{b}-\frac{m_1b}{2\pi})
}\nonumber \\
&\quad \frac{\prod_{i<j}(2\sinh\frac{\lambda_i-\lambda_j}{2b^2})^2}{\prod_{i,j}2b^2\cosh[\frac{\lambda_i-\lambda_j}{2b^2}+m_1]}.
\end{align}
Comparing this result for $m_1=m_2=-\pi i(\nu-2a)/\nu$ with $a,\nu\in\mathbb{N}$ with the fermionic spectral trace of the inverse quantum mirror curve of $Y^{\nu,0}$ geometry at the one period phases obtained in \cite{Bonelli:2017ptp}
\begin{align}
&Z_\nu(0,0,\cdots,0,\mathop{N}_a,0,0,\cdots,0,\zeta,b)\nonumber \\
&=\frac{1}{N!}\int \frac{d^Nx}{(2\pi)^N}\prod_{i=1}^N e^{\frac{\pi(\nu-a-2)b\zeta}{\nu}+\frac{(\nu-a)x_i}{\nu}}
\frac{
\Phi_b(\frac{x_i}{2\pi b}-\frac{\zeta}{2}+\frac{iab}{2\nu})
\Phi_b(\frac{x_i}{2\pi b}+\frac{\zeta}{2}+\frac{i(\nu-a)b}{2\nu})
}{
\Phi_b(\frac{x_i}{2\pi b}+\frac{\zeta}{2}-\frac{i(\nu-a)b}{2\nu})
\Phi_b(\frac{x_i}{2\pi b}-\frac{\zeta}{2}-\frac{iab}{2\nu})
}\nonumber \\
&\quad \frac{\prod_{i<j}(2\sinh\frac{x_i-x_j}{2b^2})^2}{\prod_{i,j}2b^2\cosh[\frac{x_i-x_j}{2b^2}-\frac{\pi i(\nu-2a)}{2\nu}]}
\end{align}
we find that the two results coincide with the following parameter identifications
\begin{align}
&z^NZ_{k,M}\Bigl(N,
-\frac{\pi i(\nu-2a)}{\nu},
-\frac{\pi i(\nu-2a)}{\nu}
\Bigr)\nonumber \\
&\quad =Z_{k,M}(0)(e^{\pi i(1+\frac{2}{\nu})M-\frac{\pi ik(a^2-a\nu+\nu)}{\nu^2}}z)^NZ_\nu\Bigl(0,0,\cdots,0,\mathop{N}_a,0,0,\cdots,0,-\frac{ib}{2}+\frac{iM}{b},b\Bigr)
\end{align}
Hence our conjecture \eqref{conjecturedmABJMbilinearidentity} can be rewritten as
\begin{align}
\eqref{conjecturedmABJMbilinearidentity}\quad \Leftrightarrow\quad
&\Xi_a\Bigl(-e^{\frac{2\pi i(1-a)}{\nu}}\kappa,\zeta+\frac{i}{b};b\Bigr)
\Xi_a\Bigl(-e^{-\frac{2\pi i(1-a)}{\nu}}\kappa,\zeta-\frac{i}{b};b\Bigr)
(1+e^{\frac{2\pi\zeta}{b}})\nonumber \\
&\quad =\Xi_a(\kappa,\zeta;b)^2
+e^{\frac{2\pi\zeta}{b}}
\Xi_a(e^{\frac{2\pi ia}{\nu}}\kappa,\zeta;b)
\Xi_a(e^{-\frac{2\pi ia}{\nu}}\kappa,\zeta;b),
\end{align}
with
\begin{align}
\Xi_a(\kappa,\zeta;b)=\sum_{N=0}^\infty \kappa^NZ_\nu(0,0,\cdots,0,\mathop{N}_a,0,0,\cdots,0,\zeta,b).
\end{align}
This is indeed a natural higher rank generalization of the relation between the spectral determinant of quantum mirror curve of local $\mathbb{P}^1\times \mathbb{P}^1$ and the affine $\text{SU}(2)$ $q$-Toda equation found in \cite{Bonelli:2017gdk}
More explicitly, by defining the $\tau$-function $\tau_j$ in the following way
\begin{align}
\tau_j\Bigl(w=e^{-\pi i\nu(1-\frac{2M}{k})},q=e^{\frac{2\pi i\nu}{k}};z\Bigr)=
\frac{A(e^{-\pi i\nu(1-\frac{2M}{k})},e^{\frac{2\pi i\nu}{k}})}
{Z_{k,M}(0)}
\Xi_{k,M}(e^{-\frac{2\pi iaM}{\nu}+\frac{2\pi iaj}{\nu}}z)
,\nonumber \\
A(w,q)=\prod_{m,n\ge 0}(1+(q^{-m-n-1}w)^{\frac{1}{\nu}})
\end{align}
the bilinear relation \eqref{conjecturedmABJMbilinearidentity} for $m_1=m_2=-\pi i(\nu-2a)/\nu$ reduces to the bilinear form of the affine $\text{SU}(\nu)$ $q$-Toda equation written in \cite{Bershtein:2018srt}:\footnote{
Note that in \cite{Bershtein:2018zcz} the relation between the grand partition function of the $\text{U}(N)\times \text{U}(N+M)$ ABJM theory and the partition function of the 5d ${\cal N}=1$ pure $\text{SU}(2)$ Yang-Mills theory on $S^1$ was also discussed, where the bilinear equation for the grand partition function \eqref{BGM1804} follows from the Nakajima-Yoshioka blowup equation for the 5d partition function \cite{Nakajima:2005fg}.
}
\begin{align}
\eqref{conjecturedmABJMbilinearidentity}\quad\Leftrightarrow\quad \tau_j(qw)\tau_j(q^{-1}w)&=\tau_j^2(w)+w^{\frac{1}{\nu}}\tau_{j-1}(w)\tau_{j+1}(w).
\label{BGM1804}
\end{align}

\section{Check of \eqref{conjecturedmABJMbilinearidentity} for small $z$ using open string formalism}
\label{sec_openstringformalism}

In this section we consider another way to rewrite the partition function of the mass deformed ABJM theory \eqref{ZkMNm1m2}, the variation of so called open string formalism \cite{Matsumoto:2013nya} which is useful for the computation of the exact values of the partition function.
Starting from \eqref{closed_first2} we use the following formula
\begin{align}
\frac{1}{N!}\int d^Nx\det[[f_i(x_j)]_{i,j:N\times N}]\det[[g_j(x_i)]_{i,j:N\times N}]=\det_{i,j}\Bigl[\int dxf_i(x)g_j(x)\Bigr]
\end{align}
for the integration variables $\{{\widetilde\lambda}_m\}_{m=1}^{N+M}$ to rewrite the partition function $Z_{k,M}(N,m_1,m_2)$ as
\begin{align}
&Z_{k,M}(N,m_1,m_2)\nonumber \\
&=\frac{1}{N!}\int\Big(\frac{d\lambda}{2\pi}\Bigr)^N
\det
\begin{pmatrix}
[\langle\lambda_i|
\frac{e^{-\frac{im_1{\widehat p}}{2\pi}}}{2\cosh\frac{{\widehat p}}{2}}
e^{-\frac{i}{4\pi k}{\widehat x}^2}
\frac{e^{-\frac{im_2{\widehat p}}{2\pi}}}{2\cosh\frac{{\widehat p}}{2}}
e^{\frac{i}{4\pi k}{\widehat x}^2}
|\lambda_j\rangle]_{i,j}&
[\langle\lambda_i|\frac{e^{-\frac{im_1{\widehat p}}{2\pi}}}{2\cosh\frac{{\widehat p}}{2}}e^{-\frac{i}{4\pi k}{\widehat x}^2}|2\pi i\ell_s\rangle\!\rangle]_{i,s}\\
[\langle\!\langle 2\pi i\ell_r|e^{-\frac{i}{4\pi k}{\widehat x}^2}\frac{e^{-\frac{im_2{\widehat p}}{2\pi}}}{2\cosh\frac{{\widehat p}}{2}}e^{\frac{i}{4\pi k}{\widehat x}^2}|\lambda_j\rangle]_{r,j}&
[\langle\!\langle 2\pi i\ell_r|e^{-\frac{i}{4\pi k}{\widehat x}^2}|2\pi i\ell_s\rangle\!\rangle]_{r,s}
\end{pmatrix}.
\end{align}
With this result, the grand partition function $\Xi_{k,M}(z;m_1,m_2)$ \eqref{XikM} can be written as
\begin{align}
\Xi_{k,M}(z;m_1,m_2)
&=\text{Det}\begin{pmatrix}
1+z\Bigl(
\frac{e^{-\frac{im_1{\widehat p}}{2\pi}}}{2\cosh\frac{{\widehat p}}{2}}
e^{-\frac{i}{4\pi k}{\widehat x}^2}
\frac{e^{-\frac{im_2{\widehat p}}{2\pi}}}{2\cosh\frac{{\widehat p}}{2}}
e^{\frac{i}{4\pi k}{\widehat x}^2}
\Bigr)
&
\Bigl[z\Bigl(
\frac{e^{-\frac{im_1{\widehat p}}{2\pi}}}{2\cosh\frac{{\widehat p}}{2}}e^{-\frac{i}{4\pi k}{\widehat x}^2}|2\pi i\ell_s\rangle\!\rangle\Bigr)\Bigr]_s\\
\Bigl[\langle\!\langle 2\pi i\ell_r|e^{-\frac{i}{4\pi k}{\widehat x}^2}\frac{e^{-\frac{im_2{\widehat p}}{2\pi}}}{2\cosh\frac{{\widehat p}}{2}}e^{\frac{i}{4\pi k}{\widehat x}^2}\Bigr]_r&
\Bigl[\langle\!\langle 2\pi i\ell_r|e^{-\frac{i}{4\pi k}{\widehat x}^2}|2\pi i\ell_s\rangle\!\rangle\Bigr]_{r,s}
\end{pmatrix}\nonumber \\
&=
\Xi_{k,0}(z,m_1,m_2)
\det_{r,s}[H_{r,s}(z,k,m_1,m_2)],
\label{Xi_openstringformalism}
\end{align}
where
\begin{align}
&\Xi_{k,0}(z,m_1,m_2)=
\text{Det}\Bigl[1+z\Bigl(
\frac{e^{-\frac{im_1{\widehat p}}{2\pi}}}{2\cosh\frac{{\widehat p}}{2}}
e^{-\frac{i}{4\pi k}{\widehat x}^2}
\frac{e^{-\frac{im_2{\widehat p}}{2\pi}}}{2\cosh\frac{{\widehat p}}{2}}
e^{\frac{i}{4\pi k}{\widehat x}^2}
\Bigr)\Bigr]\nonumber \\
&H_{r,s}(z,k,m_1,m_2)\nonumber \\
&=
\langle\!\langle 2\pi i\ell_r|e^{-\frac{i}{4\pi k}{\widehat x}^2}|2\pi i\ell_s\rangle\!\rangle\nonumber \\
&\quad -
\langle\!\langle 2\pi i\ell_r|e^{-\frac{i}{4\pi k}{\widehat x}^2}\frac{e^{-\frac{im_2{\widehat p}}{2\pi}}}{2\cosh\frac{{\widehat p}}{2}}e^{\frac{i}{4\pi k}{\widehat x}^2}
\frac{1}{
1+z\Bigl(
\frac{e^{-\frac{im_1{\widehat p}}{2\pi}}}{2\cosh\frac{{\widehat p}}{2}}
e^{-\frac{i}{4\pi k}{\widehat x}^2}
\frac{e^{-\frac{im_2{\widehat p}}{2\pi}}}{2\cosh\frac{{\widehat p}}{2}}
e^{\frac{i}{4\pi k}{\widehat x}^2}
\Bigr)
}
z\Bigl(
\frac{e^{-\frac{im_1{\widehat p}}{2\pi}}}{2\cosh\frac{{\widehat p}}{2}}e^{-\frac{i}{4\pi k}{\widehat x}^2}|2\pi i\ell_s\rangle\!\rangle\Bigr)\nonumber \\
&=\sum_{n=0}^\infty(-z)^n\langle\!\langle 2\pi i\ell_r|\Bigl(
e^{-\frac{i}{4\pi k}{\widehat x}^2}
\frac{e^{-\frac{im_2{\widehat p}}{2\pi}}}{2\cosh\frac{{\widehat p}}{2}}
e^{\frac{i}{4\pi k}{\widehat x}^2}
\frac{e^{-\frac{im_1{\widehat p}}{2\pi}}}{2\cosh\frac{{\widehat p}}{2}}
\Bigr)^n
e^{-\frac{i}{4\pi k}{\widehat x}^2}
|2\pi i\ell_s\rangle\!\rangle.
\end{align}
By performing similarity transformations of the 1d quantum mechanical operators we can further rewrite $\Xi_{k,0}(z,m_1,m_2)$ and $\det H_{r,s}(z)$ as
\begin{align}
\Xi_{k,0}(z,m_1,m_2)&=\text{Det}(1+z{\widehat\rho}_0),\nonumber \\
\det H_{r,s}(z)&=i^{-\frac{M}{2}}e^{-\frac{\pi iM(4M^2-1)}{6k}}\det_{r,s}\Bigl[\sum_{n=0}^\infty(-z)^n\langle\!\langle 2\pi i\ell_r|{\widehat\rho}_0^n|2\pi i\ell_s\rangle\Bigr],
\label{Openfinal}
\end{align}
with ${\widehat\rho}_0$ given as \eqref{rho0}.

\subsection{Recursive calculation of $\langle x|{\widehat\rho}_0^n|y\rangle$ and $\langle\!\langle 2\pi i\ell_r|{\widehat\rho}_0^n|x\rangle$}
\label{sec_exactcalculation}

In the previous section we have rewritten the grand partition function of the mass deformed ABJM theory $\Xi_{k,M}(z,m_1,m_2)$ as \eqref{Xi_openstringformalism}, which implies that we can calculate the partition function $Z_{k,M}(N,m_1,m_2)$, the expansion coefficients of $\Xi_{k,M}(z,m_1,m_2)$ in $z$, by calculating $\text{Tr}{\widehat\rho}_0^n$ and $\langle\!\langle 2\pi i\ell_r|{\widehat\rho}_0^n|x\rangle$.\footnote{
Although we could also perform the same calculation directly for $\text{Tr}{\widehat\rho}_M^n$ both in the method of section \ref{sec_fixingm2} and in the method of section \ref{sec_notfixingm2}, the calculation becomes more complicated as $M$ grows.
}
Though the first terms of these sequences can be obtained easily
\begin{align}
\langle\!\langle 2\pi i\ell_r|x\rangle=\frac{1}{\sqrt{k}}e^{\frac{\ell_rx}{k}},\quad
\text{Tr}{\widehat\rho}_0=\frac{1}{4k\cosh\frac{m_1}{2}\cosh\frac{m_2}{2}},\quad (k,m_1,m_2\in\mathbb{C})
\end{align}
the calculation is non-trivial for higher $n$.
Below we show two ways to calculate these quantities recursively in $n$.
The results are listed in appendix \ref{app_listofexactvalues}, with which we can check that the conjectured bilinear relation \eqref{conjecturedmABJMbilinearidentity} indeed holds for various values of $k,M$ and $m_1=m_2$.

\subsubsection{Algorithm with fixing $\frac{im_2}{2\pi}\in\mathbb{Q}$}
\label{sec_fixingm2}
To explain the first method, which was used in \cite{Nosaka:2015iiw,Honda:2018pqa} by generalizing the method used for the massless case of \cite{Tracy:1995ax,Putrov:2012zi}, first we write the matrix element $\langle x|{\widehat\rho}_0|y\rangle$ as
\begin{align}
(e^{\frac{x}{k}-\frac{m_1}{2}}+e^{\frac{y}{k}+\frac{m_1}{2}})\langle x|{\widehat\rho}_0|y\rangle=\frac{e^{-\frac{im_2}{2\pi}{\widehat x}}}{2\cosh\frac{{\widehat x}}{2}}\frac{e^{\frac{x+y}{2k}}}{k},
\end{align}
which implies
\begin{align}
e^{\frac{{\widehat x}}{k}-\frac{m_1}{2}}{\widehat\rho}_0
=
-{\widehat\rho}_0
e^{\frac{{\widehat x}}{k}+\frac{m_1}{2}}
+
\frac{e^{-\frac{im_2}{2\pi}{\widehat x}}}{2\cosh\frac{{\widehat x}}{2}}
e^{\frac{{\widehat x}}{2k}}
|0\rangle\!\rangle
\langle\!\langle 0|
e^{\frac{{\widehat x}}{2k}}.
\end{align}
Using this relation repeatedly we obtain
\begin{align}
e^{\frac{{\widehat x}}{k}-\frac{nm_1}{2}}{\widehat\rho}_0^n=(-1)^n{\widehat\rho}_0^ne^{\frac{{\widehat x}}{k}+\frac{nm_1}{2}}
+\sum_{\ell=0}^{n-1}(-1)^\ell e^{(-\frac{n-1}{2}+\ell)m_1}
{\widehat\rho}_0^\ell\frac{e^{-\frac{im_2}{2\pi}{\widehat x}}}{2\cosh\frac{{\widehat x}}{2}}e^{\frac{{\widehat x}}{2k}}|0\rangle\!\rangle
\langle\!\langle 0|
e^{\frac{{\widehat x}}{2k}}
{\widehat\rho}_0^{n-1-\ell}.
\end{align}
This implies that the matrix element of ${\widehat\rho}_0^n$ is written as
\begin{align}
\langle x|{\widehat\rho}_0^n|y\rangle=\frac{1}{e^{\frac{x}{k}-\frac{nm_1}{2}}-(-1)^ne^{\frac{y}{k}+\frac{nm_1}{2}}}\frac{e^{(\frac{1}{2k}-\frac{im_2}{2\pi})x+\frac{y}{2k}}}{2\cosh\frac{x}{2}}\sum_{\ell=0}^{n-1}(-1)^\ell\phi_\ell(x)\psi_{n-1-\ell}(y),
\label{rho0^n_TW}
\end{align}
with
\begin{align}
\phi_\ell(x)&=e^{\frac{\ell m_1}{2}}\langle x|
e^{-\frac{{\widehat x}}{2k}}\Bigl(\frac{e^{-\frac{im_2}{2\pi}{\widehat x}}}{2\cosh\frac{{\widehat x}}{2}}\Bigr)^{-1}
{\widehat\rho}_0^\ell
\frac{e^{-\frac{im_2}{2\pi}{\widehat x}}}{2\cosh\frac{{\widehat x}}{2}}
e^{\frac{{\widehat x}}{2k}}
|0\rangle\!\rangle,\nonumber \\
\psi_\ell(x)&=e^{-\frac{\ell m_1}{2}}\langle\!\langle 0|
e^{\frac{{\widehat x}}{2k}}
{\widehat\rho}_0^\ell
e^{-\frac{{\widehat x}}{2k}}
|x\rangle=\phi_\ell(x)\Bigr|_{m_1\rightarrow -m_1}.
\end{align}
The vector $\phi_\ell(x)$ obeys the following recursion relation and the initial condition
\begin{align}
\phi_0(x)&=\langle x|0\rangle\!\rangle=\frac{1}{\sqrt{k}},\nonumber \\
\phi_{\ell+1}(x)&=e^{\frac{m_1}{2}}\int\frac{dy}{2\pi}\langle x|
e^{-\frac{{\widehat x}}{2k}}
\Bigl(\frac{e^{-\frac{im_2}{2\pi}{\widehat x}}}{2\cosh\frac{{\widehat x}}{2}}\Bigr)^{-1}
{\widehat\rho}_0
\frac{e^{-\frac{im_2}{2\pi}{\widehat x}}}{2\cosh\frac{{\widehat x}}{2}}
e^{\frac{{\widehat x}}{2k}}
|y\rangle
\phi_\ell(y)\nonumber \\
&=\frac{1}{2\pi k}\int dy\frac{e^{(\frac{1}{k}+\frac{1}{2}-\frac{im_2}{2\pi})y}}{(e^{\frac{y}{k}}+e^{-m_1}e^{\frac{x}{k}})(e^y+1)}\phi_\ell(y).
\label{phirecursionrelation}
\end{align}

\begin{figure}
\begin{center}
\includegraphics[width=10cm]{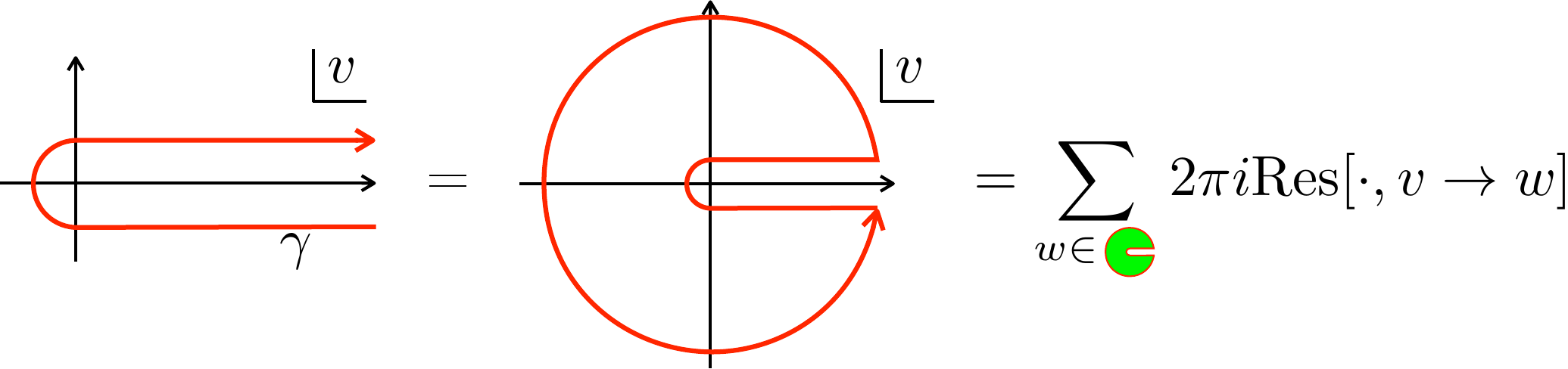}
\end{center}
\caption{
Left: The integration contour $\gamma$ used in the first line of \eqref{PYrewriting}; Center and Right: deformation of the contour to derive the second line of \eqref{PYrewriting}.
}
\label{fig_PYcontour}
\end{figure}

When $k\in\mathbb{Q}$ and $im_2/(2\pi)\in\mathbb{Q}$ we can further rewrite the recursion relation \eqref{phirecursionrelation} as follows.
By choosing $R\in\mathbb{N}$ such that the following $S,T$ are integers
\begin{align}
S=\frac{R}{k},\quad
T=R\Bigl(\frac{1}{k}+\frac{1}{2}-\frac{im_2}{2\pi}\Bigr),
\end{align}
the integration can be rewritten as ($u=e^{x/R},v=e^{y/R}$)
\begin{align}
\phi_{\ell+1}(u)=\frac{S}{2\pi}\int_0^\infty dv\frac{v^{T-1}}{(v^S+e^{-m_1}u^S)(v^R+1)}\phi_\ell(v).
\label{phirecursionrelation_in_RSTu}
\end{align}
If we further assume that $\phi_\ell(u)$ has the following structure\footnote{
This property is inductively correct in $\ell$.
}
\begin{align}
\phi_\ell(u)=\sum_{j\ge 0}\phi_\ell^{(j)}(u)(\log u)^j,\quad \phi_\ell^{(j)}(u)\text{: rational function of }u,
\end{align}
the integration \eqref{phirecursionrelation_in_RSTu} can be rewritten as \cite{Putrov:2012zi}
\begin{align}
\phi_{\ell+1}(u)
&=\frac{S}{2\pi}\sum_{j\ge 0}\biggl(-\frac{(2\pi i)^j}{j+1}\int_\gamma dv\frac{v^{T-1}}{(v^S+e^{-m_1}u^S)(v^R+1)}\phi_\ell^{(j)}(v)B_{j+1}\Bigl(\frac{\log^{(+)}v}{2\pi i}\Bigr)\biggr)\nonumber \\
&=\frac{S}{2\pi}\sum_{j\ge 0}\biggl(-\frac{(2\pi i)^{j+1}}{j+1}\sum_{w\text{: poles in }\mathbb{C}\backslash\mathbb{R}_{\ge 0}}
\text{Res}\biggl[\frac{v^{T-1}}{(v^S+e^{-m_1}u^S)(v^R+1)}\phi_\ell^{(j)}(v)B_{j+1}\Bigl(\frac{\log^{(+)}v}{2\pi i}\Bigr),v\rightarrow w\biggr]\biggr).
\label{PYrewriting}
\end{align}
where $\log^{(+)}$ is the logarithm with branch cut $\mathbb{R}_{\ge 0}$ (i.e.~$\log^{(+)} (re^{i\theta})=\log r+i\theta$ with $0<\theta\le 2\pi$), $\gamma$ is the contour depicted in figure \ref{fig_PYcontour} (left) and $B_{j+1}(z)$ is the Bernoulli polynomial.
The poles to be collected when we calculate $\phi_{\ell+1}$ from $\phi_\ell$ are the union of $\{v=e^{-m_1/S}ue^{(2\pi i/S)(a-1/2)}\}_{a=1}^{S}$, $\{v=e^{(2\pi i/R)(b-1/2)}\}_{b=1}^R$ and the poles of $\phi_\ell(v)$.
The poles of $\phi_\ell(u)$ can be listed exhaustively in the following way \cite{Nosaka:2015iiw}.
As the locations of the first set of the poles $\{v=e^{-m_1/S}ue^{(2\pi i/S)(a-1/2)}\}_{a=1}^{S}$ are $u$-dependent, the recursion relation \eqref{PYrewriting} generates new poles through the resudies at these $u$-dependent poles.
Therefore from $\{v=e^{(2\pi i/R)(b-1/2)}\}_{b=1}^R$ we find $\phi_{\ell\ge 1}(u)$ has poles at $\bigcup_{b=1}^R\{u=e^{m_1/S}e^{-(2\pi i/S)(a-1/2)}e^{(2\pi i/R)(b-1/2)}\}_{a=1}^S$.
Also, if $\phi_\ell(u)$ has a pole at $u=u_0$, it follows through the same process that $\phi_{\ell+1}(u)$ has poles at $\{u=e^{m_1/S}e^{-(2\pi i/S)(a-1/2)}u_0\}_{a=1}^S$.
Taking into account that the initial condition $\phi_0(u)$ has no poles, we conclude that the poles of $\phi_{\ell\ge 1}(u)$ are
\begin{align}
\text{poles of }\phi_\ell(u)=\bigcup_{\ell'=1}^\ell\bigcup_{a=1}^S\bigcup_{b=1}^R\Bigl\{u=e^{-\frac{\ell'm_1}{S}}e^{\frac{\pi i}{S}\text{Mod}[\ell',2]}e^{\frac{2\pi i}{S}(a-\frac{1}{2})}e^{\frac{2\pi i}{R}(b-\frac{1}{2})}\Bigr\}.
\label{polesofphiell}
\end{align}

Once we obtain $\phi_\ell(u)$ with $\ell=0,1,\cdots,n-1$ we can calculate $\text{Tr}{\widehat\rho}_0^n$ \eqref{rho0^n_TW} by the same change of integration variable and the contour deformation:
\begin{align}
\text{Tr}{\widehat\rho}^n
&=\int \frac{dx}{2\pi}\langle x|{\widehat\rho}_0^n|x\rangle\nonumber \\
&=\frac{1}{2\pi(e^{-\frac{nm_1}{2}}-(-1)^ne^{\frac{nm_1}{2}})}\int dx\frac{e^{-\frac{im_2}{2\pi}x}}{2\cosh\frac{x}{2}}\sum_{\ell=0}^{n-1}\phi_\ell(x)\psi_{n-1-\ell}(x)\nonumber \\
&=\frac{R}{2\pi(e^{-\frac{nm_1}{2}}-(-1)^ne^{\frac{nm_1}{2}})}\sum_{j\ge 0}\biggl(-\frac{(2\pi i)^{j+1}}{j+1}\sum_{w\text{: poles}}
\text{Res}\biggl[\frac{u^{T-S-1}}{u^R+1}A_n^{(j)}(u)B_{j+1}\Bigl(\frac{\log^{(+)}u}{2\pi i}\Bigr),u\rightarrow w\biggr]\biggr)
\label{trrhon}
\end{align}
where $A_n^{(j)}(u)$ are the rational functions given by
\begin{align}
\sum_{\ell=0}^{n-1}(-1)^\ell\phi_\ell(u)\psi_{n-1-\ell}(u)=\sum_{j\ge 0}A_n^{(j)}(u)(\log u)^j.
\end{align}
The poles to be collected for the calculation of $\text{Tr}{\widehat\rho}_0^n$ in \eqref{trrhon} are
\begin{align}
&
\{u=e^{(2\pi i/R)(b-1/2)}\}_{b=1}^R
\cup
\Bigl(\bigcup_{\ell=0}^{n-1}(\text{poles of }\phi_\ell(u))\Bigr)
\cup
\Bigl(\bigcup_{\ell=0}^{n-1}(\text{poles of }\phi_\ell(u)\bigr|_{m_1\rightarrow -m_1})\Bigr)\nonumber \\
&=\{u=e^{(2\pi i/R)(b-1/2)}\}_{b=1}^R
\cup
(\text{poles of }\phi_{n-1}(u))
\cup
(\text{poles of }\phi_{n-1}(u)\bigr|_{m_1\rightarrow -m_1}),
\end{align}
where the poles of $\phi_\ell(u)$ are given in \eqref{polesofphiell}.

The calculation of $\langle\!\langle 2\pi i\ell_r|{\widehat\rho}_0^n|x\rangle$ can be performed in the same way as $\phi_\ell(u)$.
Writing $\langle\!\langle 2\pi i\ell_r|{\widehat\rho}_0^n|x\rangle$ as
\begin{align}
\langle\!\langle 2\pi i\ell_r|{\widehat\rho}_0^n|x\rangle
=e^{\frac{x}{2k}}\chi_{r,n}(x),\quad
\chi_{r,n}(x)=e^{\frac{x}{2k}}\langle\!\langle 2\pi i\ell_r|e^{-\frac{{\widehat x}}{2k}}(e^{\frac{{\widehat x}}{2k}}{\widehat\rho}_0^ne^{-\frac{{\widehat x}}{2k}})|x\rangle,
\end{align}
$\chi_{r,n}(x)$ obeys the following recursion relation and the initial condition
\begin{align}
\chi_{r,0}(x)&=\frac{1}{\sqrt{k}}e^{\frac{1}{k}(\ell_r-\frac{1}{2})x},\nonumber \\
\chi_{r,n+1}(x)&=\frac{e^{\frac{m_1}{2}}}{2\pi k}\int dy\frac{e^{(\frac{1}{k}+\frac{1}{2}-\frac{im_2}{2\pi})y}}{(e^{\frac{y}{k}}+e^{m_1}e^{\frac{x}{k}})(e^y+1)}\chi_{r,n}(y).
\end{align}
The recursion relation is nothing but the one for $\psi_\ell(x)$ (except an extra overall factor $e^{m_1/2}$).
When $k\in\mathbb{Q}$ and $im_2/(2\pi)\in\mathbb{Q}$ it is rewritten as
\begin{align}
\chi_{r,0}(u)&=\frac{1}{\sqrt{k}}u^{(\ell_r-\frac{1}{2})S},\nonumber \\
\chi_{r,n+1}(u)
&=e^{\frac{m_1}{2}}\frac{S}{2\pi}\sum_{j\ge 0}\biggl(-\frac{(2\pi i)^j}{j+1}\int_\gamma dv\frac{v^{T-1}}{(v^S+e^{m_1}u^S)(v^R+1)}\chi_{r,n}^{(j)}(v)B_{j+1}\Bigl(\frac{\log^{(+)}v}{2\pi i}\Bigr)\biggr)\nonumber \\
&=e^{\frac{m_1}{2}}\frac{S}{2\pi}\sum_{j\ge 0}\biggl(-\frac{(2\pi i)^{j+1}}{j+1}\sum_{w\text{: poles}}
\text{Res}\biggl[\frac{v^{T-1}}{(v^S+e^{m_1}u^S)(v^R+1)}\chi_{r,n}^{(j)}(v)B_{j+1}\Bigl(\frac{\log^{(+)}v}{2\pi i}\Bigr),v\rightarrow w\biggr]\biggr),
\end{align}
with
\begin{align}
\chi_{r,n}(u)=\sum_{j\ge 0}\chi_{r,n}^{(j)}(u)(\log u)^j,\quad \chi_{r,n}^{(j)}(u)\text{: rational function of }u.
\end{align}
The poles to be collected for the calculation of $\chi_{r,n+1}(u)$ are
\begin{align}
\{v=e^{-m_1/S}ue^{(2\pi i/S)(a-1/2)}\}_{a=1}^{S}\cup\{v=e^{(2\pi i/R)(b-1/2)}\}_{b=1}^R\cup (\text{poles of }\chi_{r,n}(v)),
\end{align}
with $(\text{poles of }\chi_{r,n}(v))=(\text{poles of }\phi_n(v))|_{m_1\rightarrow -m_1}$.

\subsubsection{Algorithm without fixing $m_2$}
\label{sec_notfixingm2}

We can also calculate $\langle x|{\widehat\rho}_0^n|y\rangle$ and $\langle\!\langle 2\pi i\ell_r|{\widehat\rho}_0^n|x\rangle$ without fixing $m_1,m_2$ (and $\ell_r$) in the following way \cite{Okuyama:2011su,Russo:2015exa}.
Since these quantities are meromorphic functions in $m_1,m_2$, we assume $m_1,m_2\in\mathbb{R}$ in the calculation and the results for $m_1,m_2\in\mathbb{C}$ are obtained by the analytic continuation.
The algorithm works for general $k\in\mathbb{Q}$, which we denote $k=p/q$ with $p,q$ coprime.

\begin{figure}
\begin{center}
\includegraphics[width=12cm]{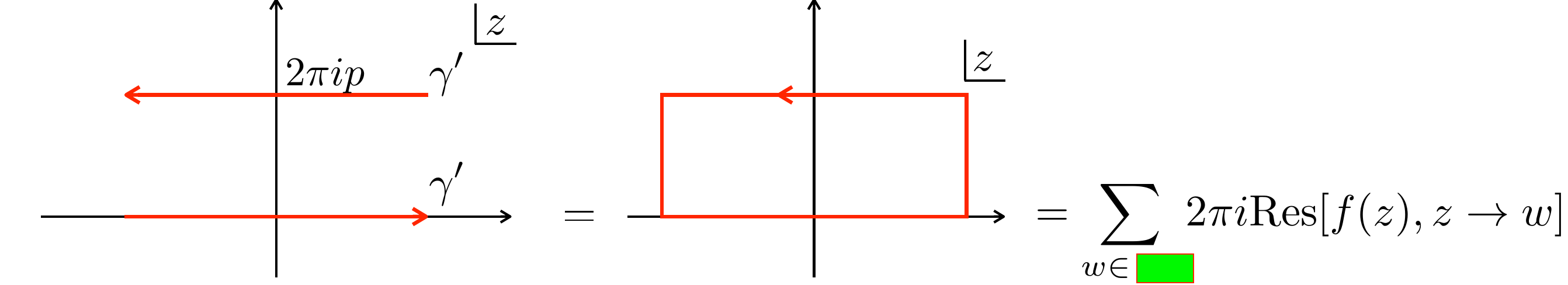}
\end{center}
\caption{Contour $\gamma'$ used in \eqref{Okuyamamethodresidueintegral} and its deformation.}
\label{fig_contour_okuyamamethod}
\end{figure}

First we consider $\langle\!\langle 2\pi i\ell_r|{\widehat\rho}^n|x\rangle$ which obeys the following recursion relation
\begin{align}
\langle\!\langle 2\pi i\ell_r|x\rangle&=\frac{1}{\sqrt{k}}e^{\frac{\ell_rx}{k}},\nonumber \\
\langle\!\langle 2\pi i\ell_r|{\widehat\rho}_0|x\rangle&=\int \frac{dy}{2\pi} \langle\!\langle 2\pi i\ell_r|y\rangle\langle y|{\widehat\rho}_0|x\rangle
=\frac{1}{8\pi k\sqrt{k}}\int dy\frac{e^{(\frac{\ell_r}{k}-\frac{im_2}{2\pi})y}}{\cosh\frac{y}{2}\cosh\frac{y-x-m_1k}{2k}},\label{<<2piiellr|rho0|x>} \\
\langle\!\langle 2\pi i\ell_r|{\widehat\rho}_0^{n+1}|x\rangle&=\int \frac{dy}{2\pi} \langle\!\langle 2\pi i\ell_r|{\widehat\rho}_0^n|y\rangle\langle y|{\widehat\rho}_0|x\rangle.
\label{recursion_Okuyama}
\end{align}
To perform the integration \eqref{recursion_Okuyama}, let us consider the following integration
\begin{align}
I[f]=\int_{-\infty}^\infty dzf(z).
\label{If}
\end{align}
If we assume that $f(x)$ is quasi-periodic as $f(x+2\pi ip)=Af(x)$ with some constant $A$ (this is true for the integrand of $\langle\!\langle 2\pi i\ell_r|{\widehat\rho}_0|x\rangle$ \eqref{<<2piiellr|rho0|x>}, with $A=(-1)^{p+q}e^{(\frac{\ell_r}{k}-\frac{im_2}{2\pi})2\pi ip}$), then we find that the integration \eqref{If} is invariant, up to an overall factor, under the shift of the integration contour
\begin{align}
\int_{-\infty+2\pi ip}^{\infty+2\pi ip} dzf(z)
=\int_{-\infty}^{\infty} dzf(z+2\pi ip)
=A \int_{-\infty}^\infty dzf(z).
\end{align}
Hence the integration $I[f]$ can be evaluated as
\begin{align}
I[f]
=\frac{1}{1-A}\int_{\gamma'} dzf(z)
=\frac{2\pi i}{1-A}\sum_{w}\text{Res}\Bigl[f(z),z\rightarrow w\Bigr],
\label{Okuyamamethodresidueintegral}
\end{align}
where the integration contour $\gamma'$ is as displayed in figure \ref{fig_contour_okuyamamethod}.
For example, applying this calculation to $\langle\!\langle 2\pi i\ell_r|{\widehat\rho}_0|x\rangle$ \eqref{<<2piiellr|rho0|x>} we obtain
\begin{align}
\langle\!\langle 2\pi i\ell_r|{\widehat\rho}_0|x\rangle&=
\frac{1}{8 \pi k\sqrt{k}}\frac{2\pi i}{1-(-1)^{p+q}e^{(\frac{\ell_r}{k}-\frac{im_2}{2\pi})2\pi ip}}\biggl[
\sum_{n=1}^p\frac{2i(-1)^ne^{(\frac{\ell_r}{k}-\frac{im_2}{2\pi})2\pi i(n-\frac{1}{2})}}{\cosh[\frac{1}{2k}(2\pi i(n-\frac{1}{2})-x-m_1k)]}\nonumber \\
&\quad +\sum_{n=1}^q\frac{2ik(-1)^ne^{(\frac{\ell_r}{k}-\frac{im_2}{2\pi})(x+m_1k+2\pi ik(n-\frac{1}{2})}}{\cosh[\frac{1}{2}(x+m_1k+2\pi ik(n-\frac{1}{2}))]}
\biggr].
\label{<<2piiellr|rho0|x>result}
\end{align}
Next we consider $\langle\!\langle 2\pi i\ell_r|{\widehat\rho}_0^2|x\rangle$ which is given as \eqref{recursion_Okuyama}
\begin{align}
\langle\!\langle 2\pi i\ell_r|{\widehat\rho}_0^2|x\rangle&=\int \frac{dy}{2\pi} \langle\!\langle 2\pi i\ell_r|y\rangle\langle y|{\widehat\rho}_0|x\rangle
=\frac{1}{4\pi k}\int dy\langle\!\langle 2\pi i\ell_r|{\widehat\rho}_0|y\rangle\frac{e^{-\frac{im_2}{2\pi}y}}{\cosh\frac{y}{2}\cosh\frac{y-x-m_1k}{2k}}.
\end{align}
Note that the integrand does not enjoy the quasi-periodicity in $y$ any more.
Nevertheless, if we look at the contributions from each pole \eqref{<<2piiellr|rho0|x>result}, they separately enjoy the quasi-periodicity, hence we can repeat the same calculation as above.
In principle we can repeat this algorithm indefinitely for higher $n$ by applying the formula \eqref{Okuyamamethodresidueintegral} to each term in $\langle \!\langle 2\pi i\ell_r|{\widehat\rho}_0^{n-1}|y\rangle\langle y|{\widehat\rho}_0|x\rangle$ which takes the following form
\begin{align}
\alpha e^{\beta x}
\prod_{\delta_a^{(1)}\in\Delta^{(1)}}
\frac{1}{\cosh\frac{x-\delta^{(1)}_a}{2}}
\prod_{\delta_b^{(2)}\in\Delta^{(2)}}
\frac{1}{\cosh\frac{x-\delta^{(2)}_b}{2k}},
\label{alphabetaDelta1Delta2}
\end{align}
with $\alpha,\beta$ being some constants and $\Delta^{(1)}=\{\delta_a^{(1)}\}_a,\Delta^{(2)}=\{\delta_b^{(2)}\}_b$ being some sets of constants.
Explicitly, the multiplication of ${\widehat\rho}_0$ to $\langle\!\langle 2\pi i\ell_r|{\widehat\rho}_0^{n-1}$ from the right transforms each term of the form \eqref{alphabetaDelta1Delta2} as follows
\begin{align}
&\int \frac{dy}{2\pi}\alpha e^{-\frac{im_2\beta}{2\pi}y}
\prod_{\delta_a^{(1)}\in\Delta^{(1)}}
\frac{1}{\cosh\frac{y-\delta^{(1)}_a}{2}}
\prod_{\delta_b^{(2)}\in\Delta^{(2)}}
\frac{1}{\cosh\frac{y-\delta^{(2)}_b}{2k}}
\frac{e^{-\frac{im_2}{2\pi}y}}{4k\cosh\frac{y}{2}\cosh\frac{y-x-m_1k}{2k}}\nonumber \\
&=\sum_{(\alpha',\beta',\Delta^{'(1)},\Delta^{'(2)})}
\alpha' e^{\beta' x}
\prod_{\delta_a^{(1)}\in\Delta^{'(1)}}
\frac{1}{\cosh\frac{x-\delta^{(1)}_a}{2}}
\prod_{\delta_b^{(2)}\in\Delta^{'(2)}}
\frac{1}{\cosh\frac{x-\delta^{(2)}_b}{2k}},
\end{align}
where $(\alpha',\beta',\Delta^{'(1)},\Delta^{'(2)})$ runs over all of the following (i),(ii),(iii)
\begin{itemize}
\item [(i)]
$(\alpha',\beta',\Delta^{'(1)},\Delta^{'(2)})$ for each $\delta_a^{(1)}\in \Delta^{(1)}\cup \{0\}$ and $j\in I(\delta_a^{(1)})\cup \bar{I}(\delta_a^{(1)})$ given as
\begin{align}
\alpha'&=
\begin{cases}
\displaystyle \alpha_0\frac{2i(-1)^je^{(\beta-\frac{im_2}{2\pi})(\delta_a^{(1)}+2\pi i(j-\frac{1}{2}))}}{\prod_{\substack{\delta'\in\Delta^{(1)}\cup \{0\}\\ (\delta'\neq \delta_a^{(1)})}}\cosh\frac{\delta_a^{(1)}+2\pi i(j-\frac{1}{2})-\delta'}{2}\prod_{\delta'\in\Delta^{(2)}}\cosh\frac{\delta_a^{(1)}+2\pi i(j-\frac{1}{2})-\delta'}{2k}},\quad (j\in I(\delta_a^{(1)}))\\
\displaystyle \frac{\alpha_0}{2}\frac{2i(-1)^je^{(\beta-\frac{im_2}{2\pi})(\delta_a^{(1)}+2\pi i(j-\frac{1}{2}))}}{\prod_{\substack{\delta'\in\Delta^{(1)}\cup \{0\}\\ (\delta'\neq \delta_a^{(1)})}}\cosh\frac{\delta_a^{(1)}+2\pi i(j-\frac{1}{2})-\delta'}{2}\prod_{\delta'\in\Delta^{(2)}}\cosh\frac{\delta_a^{(1)}+2\pi i(j-\frac{1}{2})-\delta'}{2k}},\quad (j\in \bar{I}(\delta_a^{(1)}))\\
\end{cases},\nonumber \\
\beta'&=0,\nonumber \\
\Delta^{'(1)}&=\{\},\nonumber \\
\Delta^{'(2)}&=\Bigl\{\delta_a^{(1)}+2\pi i\Bigl(j-\frac{1}{2}\Bigr)-m_1k\Bigr\},
\end{align}
\item [(ii)] $(\alpha',\beta',\Delta^{'(1)},\Delta^{'(2)})$ for each $\delta_b^{(2)}\in \Delta^{(2)}$ and $j\in J(\delta_b^{(2)})\cup \bar{J}(\delta_b^{(2)})$ given as
\begin{align}
\alpha'&=
\begin{cases}
\displaystyle \alpha_0\frac{2ik(-1)^je^{(\beta-\frac{im_2}{2\pi})(\delta_b^{(2)}+2\pi ik(j-\frac{1}{2}))}}{\prod_{\delta'\in\Delta^{(1)}\cup \{0\}}\cosh\frac{\delta_b^{(2)}+2\pi ik(j-\frac{1}{2})-\delta'}{2}\prod_{\substack{\delta'\in\Delta^{(2)}\\ (\delta'\neq \delta_b^{(2)})}}\cosh\frac{\delta_b^{(2)}+2\pi ik(j-\frac{1}{2})-\delta'}{2k}},\quad (j\in J(\delta_b^{(2)}))\\
\displaystyle \frac{\alpha_0}{2}\frac{2ik(-1)^je^{(\beta-\frac{im_2}{2\pi})(\delta_b^{(2)}+2\pi ik(j-\frac{1}{2}))}}{\prod_{\delta'\in\Delta^{(1)}\cup \{0\}}\cosh\frac{\delta_b^{(2)}+2\pi ik(j-\frac{1}{2})-\delta'}{2}\prod_{\substack{\delta'\in\Delta^{(2)}\\ (\delta'\neq \delta_b^{(2)})}}\cosh\frac{\delta_b^{(2)}+2\pi ik(j-\frac{1}{2})-\delta'}{2k}},\quad (j\in \bar{J}(\delta_b^{(2)}))\\
\end{cases},\nonumber \\
\beta'&=0,\nonumber \\
\Delta^{'(1)}&=\{\},\nonumber \\
\Delta^{'(2)}&=\Bigl\{\delta_b^{(2)}+2\pi ik\Bigl(j-\frac{1}{2}\Bigr)-m_1k\Bigr\},
\end{align}
\item [(iii)] and $(\alpha',\beta',\Delta^{'(1)},\Delta^{'(2)})$ for each $j=1,2,\cdots,q$ given as
\begin{align}
\alpha &=\alpha_0\cdot 2ik(-1)^j e^{(\beta-\frac{im_2}{2\pi})(m_1k+2\pi ik(j-\frac{1}{2}))},\nonumber \\
\beta &=\beta-\frac{im_2}{2\pi},\nonumber \\
\Delta^{'(1)}&=\Bigl\{\delta_a^{(1)}-m_1k-2\pi ik\Bigl(j-\frac{1}{2}\Bigr)\Bigr\}_{\delta_a^{(1)}\in\Delta^{(1)}\cup \{0\}},\nonumber \\
\Delta^{'(2)}&=\Bigl\{\delta_b^{(2)}-m_1k-2\pi ik\Bigl(j-\frac{1}{2}\Bigr)\Bigr\}_{\delta_b^{(2)}\in\Delta^{(2)}},
\end{align}
\end{itemize}
with
\begin{align}
\alpha_0&=\frac{\alpha}{8\pi k}\frac{2\pi i}{1-(-1)^{p(|\Delta^{(1)}|+1)+q(|\Delta^{(2)}|+1)}e^{(\beta-\frac{im_2}{2\pi})2\pi ip}},\nonumber \\
I(\delta_a^{(1)})&=\Bigl\{j\in\mathbb{Z}\,|\,0<\text{Im}[\delta_a^{(1)}]+2\pi\Bigl(j-\frac{1}{2}\Bigr)<2\pi p\Bigr\},\nonumber \\
\bar{I}(\delta_a^{(1)})&=\Bigl\{j\in\mathbb{Z}\,|\,\text{Im}[\delta_a^{(1)}]+2\pi\Bigl(j-\frac{1}{2}\Bigr)=0,2\pi p\Bigr\},\nonumber \\
J(\delta_b^{(2)})&=\Bigl\{j\in\mathbb{Z}\,|\,0<\text{Im}[\delta_b^{(2)}]+2\pi k\Bigl(j-\frac{1}{2}\Bigr)<2\pi p\Bigr\},\nonumber \\
\bar{J}(\delta_b^{(2)})&=\Bigl\{j\in\mathbb{Z}\,|\,\text{Im}[\delta_b^{(2)}]+2\pi k\Bigl(j-\frac{1}{2}\Bigr)=0,2\pi p\Bigr\},
\label{IIbarJJbar}
\end{align}
where $|\Delta^{(1)}|,|\Delta^{(2)}|$ are the length of $\Delta^{(1)},\Delta^{(2)}$.
Here the elements of $I,\bar{I}$ label the poles of the first cosine factor in \eqref{alphabetaDelta1Delta2} and the elements of $J,\bar{J}$ label the poles of the second cosine factors in \eqref{alphabetaDelta1Delta2}.
Note that the integrations are well defined by the principal value prescription when the poles are on top of the integration contour $\gamma'$ depicted in figure \ref{fig_contour_okuyamamethod}.
As a result, these poles contributes in the integration \eqref{Okuyamamethodresidueintegral} with the residue weighted by $1/2$.
We have counted these poles separately from $I(\delta_a^{(1)}),J(\delta_b^{(2)})$ as $\bar{I}(\delta_a^{(1)}),\bar{J}(\delta_b^{(2)})$.

The same integration technique is also applicable to the calculation of $\text{Tr}{\widehat\rho}^n$ through the decomposition \eqref{rho0^n_TW}, which we can also rewrite as
\begin{align}
\text{Tr}{\widehat\rho}_0^n&=\frac{1}{2\pi(e^{-\frac{nm_1}{2}}-(-1)^me^{\frac{nm_1}{2}})}\sum_{\ell=0}^{n-1}(-1)^\ell e^{-\frac{(n-1-2\ell)m_1}{2\pi}}\nonumber \\
&\quad \int dx \frac{e^{(-\frac{1}{k}-\frac{im_2}{2\pi})x}}{2\cosh\frac{x}{2}}\Bigl[\langle\!\langle \pi i|{\widehat\rho}_0^\ell|x\rangle\Bigr]_{m_1\rightarrow -m_1}\langle\!\langle \pi i|{\widehat\rho}_0^{n-1-\ell}|x\rangle.
\end{align}
The integrand in the last line is a linear combination of the terms of the form \eqref{alphabetaDelta1Delta2}, hence the integration can be evaluated with the following formula
\begin{align}
&\int dx e^{\beta x}\prod_{\delta_a^{(1)}\in\Delta^{(1)}}\frac{1}{\cosh\frac{x-\delta_a^{(1)}}{2}}\prod_{\delta_b^{(2)}\in\Delta^{(2)}}\frac{1}{\cosh\frac{x-\delta_b^{(2)}}{2k}}\nonumber \\
&=\frac{2\pi i}{1-(-1)^{p|\Delta^{(1)}|+q|\Delta^{(2)}|}e^{2\pi i\beta p}}
\Biggl[\nonumber \\
&\sum_{\delta_a^{(1)}\in\Delta^{(1)}}
\Bigl(
\sum_{j\in I(\delta_a^{(1)})}+\frac{1}{2}\sum_{j\in \bar{I}(\delta_a^{(1)})}\Bigr)
\frac{2i(-1)^je^{\beta(\delta_a^{(1)}+2\pi i(j-\frac{1}{2}))}}{
\prod_{\substack{\delta'\in\Delta^{(1)}\\ (\delta'\neq\delta_a^{(1)})}}\cosh\frac{\delta_a^{(1)}+2\pi i(j-\frac{1}{2})-\delta'}{2}\prod_{\delta'\in\Delta^{(2)}}\cosh\frac{\delta_a^{(1)}+2\pi i(j-\frac{1}{2})-\delta'}{2k}}\nonumber \\
&+\sum_{\delta_b^{(2)}\in\Delta^{(2)}}
\Bigl(
\sum_{j\in J(\delta_b^{(2)})}+\frac{1}{2}\sum_{j\in \bar{J}(\delta_b^{(2)})}\Bigr)
\frac{2ik(-1)^je^{\beta(\delta_b^{(2)}+2\pi ik(j-\frac{1}{2}))}}{
\prod_{\delta'\in\Delta^{(1)}}\cosh\frac{\delta_b^{(2)}+2\pi ik(j-\frac{1}{2})-\delta'}{2}\prod_{\substack{\delta'\in\Delta^{(2)}\\ (\delta'\neq \delta_b^{(2)})}}\cosh\frac{\delta_b^{(2)}+2\pi ik(j-\frac{1}{2})-\delta'}{2k}}
\Biggr],
\end{align}
with $I,\bar{I},J,\bar{J}$ defined as \eqref{IIbarJJbar}.

We have displayed the results for $n\le 2$ in appendix \ref{app_listofexactvalueswithoutfixing}.
Though the exact expressions for $\langle\!\langle 2\pi i\ell_r|{\widehat\rho}_0^n|x\rangle$ and $\text{Tr}{\widehat\rho}_0^n$ with $n>2$ are too lengthy to display, we can proceed by using Mathematica for higher $n$.
We have performed the computation for $n\le n_\text{max}$ with $n_\text{max}=7$ for $k=2$, $n_\text{max}=6$ for $k=3$, $n_\text{max}=5$ for $k=2/3,3/2,4,5,6$ and $n_\text{max}=4$ for several other $k\in\mathbb{Q}$.
Plugging them into the grand partition function $\Xi_{k,M}(z,m_1,m_2)$ and substituting a generic complex irrational value to $m_1=m_2$ (see table \ref{exactvaluesforhighernk23} for an example) we have checked that the conjectured difference identity \eqref{conjecturedmABJMbilinearidentity} indeed holds up to ${\cal O}(z^{n_\text{max}+1})$.\footnote{
One can generate the exact expressions before the substitution of $m_1,m_2$ (and $\ell_r$) by using the Mathematica notebook (\url{201130_ZkMNanalytic_withoutfixingm1andm2_forarXiv.nb}) attached to the source of this paper available at arXiv.org.
}
\begin{table}
\begin{center}
\begin{tabular}{|r|l|}
\hline
$Z_{2,0}(3)=$
$+ 0.0000034444420172682385395798807365420 - 0.0000337240454862038672844932152635140 i$\\
$Z_{2,1}(3)=$
$+ 0.0000382099395880497741652506265650111 - 0.0001306988554007833108050817131583761 i$\\
$Z_{2,2}(3)=$
$+ 0.0055770715242095177114384716730471315 - 0.0051594174158106615967121403931445771 i$\\
$Z_{2,0}(4)=$
$- 0.0000000775897470159198305637599547446 - 0.0000001011756518921727471059758984641 i$\\
$Z_{2,1}(4)=$
$+ 0.0000009438711663593511086188163930089 + 0.0000004087405746241788655917811738612 i$\\
$Z_{2,2}(4)=$
$- 0.0001653504285917934939286704610670769 - 0.0000528040240442483985487431414015146 i$\\
$Z_{2,0}(5)=$
$- 0.0000000002198192586914478736409798007 - 0.0000000000144573441669505148954356116 i$\\
$Z_{2,1}(5)=$
$- 0.0000000015615260083466536155975551341 + 0.0000000033054672575793205907416740374 i$\\
$Z_{2,2}(5)=$
$+ 0.0000005400191436028444411690234941665 + 0.0000017398055036685492156632905623505 i$\\
$Z_{2,0}(6)=$
$- 0.0000000000001225246907691902070000592 + 0.0000000000001453383429551384996297813 i$\\
$Z_{2,1}(6)=$
$- 0.0000000000061037631599002379463797156 - 0.0000000000025161595627951093002933793 i$\\
$Z_{2,2}(6)=$
$+ 0.0000000051898016505482834873280920880 - 0.0000000080148960702476791359306225233 i$\\
$Z_{2,0}(7)=$
$+ 0.0000000000000000268721667025249750661 + 0.0000000000000000828755420390328416172 i$\\
$Z_{2,1}(7)=$
$+ 0.0000000000000016760647686768458500438 - 0.0000000000000061933585687191762144848 i$\\
$Z_{2,2}(7)=$
$- 0.0000000000260602425659273907179156714 + 0.0000000000052426275884208123545727589 i$\\ \hline
$Z_{3,0}(3)=$
$- 0.0000003624144197490739076461212007997 - 0.0000035728856599566174916776594878796 i$\\
$Z_{3,1}(3)=$
$- 0.0000001940966004085613459049092799252 - 0.0000060497222546970602430145904606661 i$\\
$Z_{3,2}(3)=$
$+ 0.0000718812214806444033399076526720526 - 0.0000613657071623774821546578412393692 i$\\
$Z_{3,3}(3)=$
$+ 0.0120243169893872814065687444287212973 - 0.0007771609148023198010871758084355482 i$\\
$Z_{3,0}(4)=$
$- 0.0000000032544047194544247743023756588 - 0.0000000020599834540448572324793551042 i$\\
$Z_{3,1}(4)=$
$+ 0.0000000114317197421452279343587437785 - 0.0000000008397033914139095727378915180 i$\\
$Z_{3,2}(4)=$
$- 0.0000004361592923084441504010568557759 - 0.0000000689643523739477940325418474577 i$\\
$Z_{3,3}(4)=$
$+ 0.0001904527781369864160904152841709314 + 0.0000340107192713545756876039885790790 i$\\
$Z_{3,0}(5)=$
$- 0.0000000000014252023675918460374692040 + 0.0000000000007058594274230634453716236 i$\\
$Z_{3,1}(5)=$
$+ 0.0000000000019660390582181447625458308 + 0.0000000000084660671165274620805138313 i$\\
$Z_{3,2}(5)=$
$+ 0.0000000005084964943090853649315955503 + 0.0000000006502428131574895212907640101 i$\\
$Z_{3,3}(5)=$
$+ 0.0000011428358881531566757794631500549 + 0.0000003525014720140771192656697232273 i$\\
$Z_{3,0}(6)=$
$+ 0.0000000000000000054084266859084506734 + 0.0000000000000002802794650757403098217 i$\\
$Z_{3,1}(6)=$
$- 0.0000000000000025585161793699259602065 + 0.0000000000000013503932352293413300482 i$\\
$Z_{3,2}(6)=$
$- 0.0000000000000106248072872002466459743 - 0.0000000000006771417343271320006643413 i$\\
$Z_{3,3}(6)=$
$+ 0.0000000030000077098180098252306104505 + 0.0000000009795347258697361512361540985 i$\\ \hline
\end{tabular}
\end{center}
\caption{
Exact values of $Z_{k,M}(N,m_1,m_2)$ for a generic complex value of $m_1=m_2$ which we have chosen as $m_1=m_2=(\sqrt{2}+\sqrt{3}i)/2$.
Together with the exact values of $Z_{k,M}(N,m_1,m_2)$ for $N=1,2$ in appendix \ref{app_listofexactvalueswithoutfixing}, the displayed digits of the numerical values of $Z_{k,M}(N,m_1,m_2)$ for $N\ge 3$ are sufficient for checking the bilinear relation \eqref{conjecturedmABJMbilinearidentity} at order $\sim 10^{-35}$ which is smaller than the values of $Z_{k,M}(N,m_1,m_2)$ themselves.
}
\label{exactvaluesforhighernk23}
\end{table}
\begin{table}
\begin{center}
\begin{tabular}{|r|l|}
\hline
$Z_{4,0}(3)=$
$- 0.0000001596445009149086866982190156886 - 0.0000005893754557740624422759035272541 i$\\
$Z_{4,1}(3)=$
$- 0.0000001465567737466335850362187127074 - 0.0000005052411444397839820815223392263 i$\\
$Z_{4,2}(3)=$
$+ 0.0000021297854747687939618273009442509 - 0.0000022786180623580342900511214509227 i$\\
$Z_{4,3}(3)=$
$+ 0.0001155417403895295862777530358554630 + 0.0000581461100189894947263730010540226 i$\\
$Z_{4,4}(3)=$
$+ 0.0116701618018141557201394071902443487 + 0.0283643205840862577797331359782158014 i$\\
$Z_{4,0}(4)=$
$- 0.0000000002245870767997434170393719570 - 0.0000000000638120085896578696576695863 i$\\
$Z_{4,1}(4)=$
$+ 0.0000000002875651177599830256526624701 - 0.0000000001428393332266807229488861246 i$\\
$Z_{4,2}(4)=$
$- 0.0000000040596022308156277167049398712 + 0.0000000004564191665086604767092843196 i$\\
$Z_{4,3}(4)=$
$+ 0.0000004159888885917629489285579166084 + 0.0000002416440391255739738835924041570 i$\\
$Z_{4,4}(4)=$
$- 0.0003489532990006297368756008335046954 - 0.0002558696412186837064061226844274205 i$\\
$Z_{4,0}(5)=$
$- 0.0000000000000185361279593027355181683 + 0.0000000000000229856087210107938159293 i$\\
$Z_{4,1}(5)=$
$+ 0.0000000000000476042209753415980590695 + 0.0000000000000488828659066468711588651 i$\\
$Z_{4,2}(5)=$
$+ 0.0000000000016888416635275940690630909 + 0.0000000000008491293728894618175042306 i$\\
$Z_{4,3}(5)=$
$+ 0.0000000005586786092183451398692303713 + 0.0000000002729603154846334571280718182 i$\\
$Z_{4,4}(5)=$
$+ 0.0000020161976172384442584214274868772 - 0.0000001102502542289031043734679611258 i$\\ \hline
$Z_{5,0}(3)=$
$- 0.0000000538625413287777439633839437026 - 0.0000001281089497721297597582235865907 i$\\
$Z_{5,1}(3)=$
$- 0.0000000321427810923591851335759476735 - 0.0000000606385523646245934882290279881 i$\\
$Z_{5,2}(3)=$
$+ 0.0000001054299649109944217978475462033 - 0.0000001526552356050722035920051547306 i$\\
$Z_{5,3}(3)=$
$+ 0.0000023320730371956212019514911064430 + 0.0000018610142153383560982821863284218 i$\\
$Z_{5,4}(3)=$
$- 0.0001585068672956752422494115548000845 + 0.0002135561951970203406798693839909184 i$\\
$Z_{5,5}(3)=$
$- 0.1031259742500562377075939448189562446 + 0.0169135517524590855900487880467794458 i$\\
$Z_{5,0}(4)=$
$- 0.0000000000218467249584324831444784896 - 0.0000000000008041865823915482384782956 i$\\
$Z_{5,1}(4)=$
$+ 0.0000000000111875420459892425591996463 - 0.0000000000111723004607663049088415939 i$\\
$Z_{5,2}(4)=$
$- 0.0000000000759496509405361365876994969 + 0.0000000000314952132068155609175299774 i$\\
$Z_{5,3}(4)=$
$+ 0.0000000027354044065362251696623063658 + 0.0000000017927990586668796068172405500 i$\\
$Z_{5,4}(4)=$
$- 0.0000000789784616504830649623848447496 - 0.0000009199022388064039967710790197935 i$\\
$Z_{5,5}(4)=$
$- 0.0001327526369363546108485177212160229 + 0.0014890277071277731403881595972723748 i$\\
$Z_{5,0}(5)=$
$- 0.0000000000000003057542357421243675205 + 0.0000000000000009445915970088848706607 i$\\
$Z_{5,1}(5)=$
$+ 0.0000000000000010548737730173241160021 + 0.0000000000000003724121013099469111795 i$\\
$Z_{5,2}(5)=$
$+ 0.0000000000000114832614530517613964406 + 0.0000000000000004142818726145883299348 i$\\
$Z_{5,3}(5)=$
$+ 0.0000000000010626921634047207137704265 + 0.0000000000003795214332796580112323956 i$\\
$Z_{5,4}(5)=$
$+ 0.0000000008029802739792686237250302548 + 0.0000000005636755082281322268136100023 i$\\
$Z_{5,5}(5)=$
$+ 0.0000060720706481889983498408540301161 + 0.0000011165583251916960037326425189206 i$\\ \hline
\end{tabular}
\end{center}
\caption{
Exact values of $Z_{k,M}(N,m_1,m_2)$ for a generic complex value of $m_1=m_2$ which we have chosen as $m_1=m_2=(\sqrt{2}+\sqrt{3}i)/2$.
Together with the exact values of $Z_{k,M}(N,m_1,m_2)$ for $N=1,2$ in appendix \ref{app_listofexactvalueswithoutfixing}, the displayed digits of the numerical values of $Z_{k,M}(N,m_1,m_2)$ for $N\ge 3$ are sufficient for checking the bilinear relation \eqref{conjecturedmABJMbilinearidentity} at order $\sim 10^{-35}$ which is smaller than the values of $Z_{k,M}(N,m_1,m_2)$ themselves.
}
\label{exactvaluesforhighernk45}
\end{table}
\begin{table}
\begin{center}
\begin{tabular}{|r|l|}
\hline
$Z_{6,0}(3)=$
$- 0.0000000188552626197705052571074987233 - 0.0000000338401370969072059866884900532 i$\\
$Z_{6,1}(3)=$
$- 0.0000000072482340441164305163800784622 - 0.0000000093594423868278760651176075790 i$\\
$Z_{6,2}(3)=$
$+ 0.0000000072320414195529983402543733231 - 0.0000000147916447449864723188188515761 i$\\
$Z_{6,3}(3)=$
$+ 0.0000000868445304639474070144098687386 + 0.0000000791903995882871540607245576711 i$\\
$Z_{6,4}(3)=$
$- 0.0000040512839031169591125997384833314 + 0.0000014525922292652764348565975468802 i$\\
$Z_{6,5}(3)=$
$- 0.0001452106841333720748498550817956855 - 0.0007071062081518852152195075771069289 i$\\
$Z_{6,6}(3)=$
$+ 0.2505254348678724153644915439688414008 - 0.3569655165583691193053947443377582895 i$\\
$Z_{6,0}(4)=$
$- 0.0000000000027387887443163394257723892 + 0.0000000000004611866521191426865762459 i$\\
$Z_{6,1}(4)=$
$+ 0.0000000000005618528091890240720302622 - 0.0000000000010045912002074928039342322 i$\\
$Z_{6,2}(4)=$
$- 0.0000000000022182480127013892637886726 + 0.0000000000017243523490919009395982825 i$\\
$Z_{6,3}(4)=$
$+ 0.0000000000387395572622789905734941030 + 0.0000000000217587737892557134095941611 i$\\
$Z_{6,4}(4)=$
$+ 0.0000000015402164805127717313043091631 - 0.0000000042320725193635706934079362001 i$\\
$Z_{6,5}(4)=$
$- 0.0000025978094878185641913656163819865 + 0.0000003275390333328004741570211943445 i$\\
$Z_{6,6}(4)=$
$+ 0.0069687367541545497982908774232373760 + 0.0007355730032238962563469709620923570 i$\\
$Z_{6,0}(5)=$
$+ 0.0000000000000000000879700106831326479 + 0.0000000000000000507480212036513841242 i$\\
$Z_{6,1}(5)=$
$+ 0.0000000000000000305872509274290405211 - 0.0000000000000000017494922782929244455 i$\\
$Z_{6,2}(5)=$
$+ 0.0000000000000001272083900029037914798 - 0.0000000000000000483705188554789463527 i$\\
$Z_{6,3}(5)=$
$+ 0.0000000000000047728643339697233439353 + 0.0000000000000006192448266768018586371 i$\\
$Z_{6,4}(5)=$
$+ 0.0000000000008438128108008514841707056 + 0.0000000000010303530135401057545733039 i$\\
$Z_{6,5}(5)=$
$+ 0.0000000005482179698263412525780693423 + 0.0000000025277154129660903777324105713 i$\\
$Z_{6,6}(5)=$
$+ 0.0000156795115141421507588569121456007 + 0.0000240344345268537578114640288334317 i$\\ \hline
\end{tabular}
\end{center}
\caption{
Exact values of $Z_{k,M}(N,m_1,m_2)$ for a generic complex value of $m_1=m_2$ which we have chosen as $m_1=m_2=(\sqrt{2}+\sqrt{3}i)/2$.
Together with the exact values of $Z_{k,M}(N,m_1,m_2)$ for $N=1,2$ in appendix \ref{app_listofexactvalueswithoutfixing}, the displayed digits of the numerical values of $Z_{k,M}(N,m_1,m_2)$ for $N\ge 3$ are sufficient for checking the bilinear relation \eqref{conjecturedmABJMbilinearidentity} at order $\sim 10^{-35}$ which is smaller than the values of $Z_{k,M}(N,m_1,m_2)$ themselves.
}
\label{exactvaluesforhighernk6}
\end{table}

\section{Discussion}
\label{sec_discuss}

In this paper we have studied the $S^3$ partition function of the mass deformed $\text{U}(N)_k\times \text{U}(N+M)_{-k}$ ABJM theory for finite $k,N,M$.
For $M=0$ this model was studied by the Fermi gas formalism in \cite{Nosaka:2015iiw}.
In this paper we have extended the Fermi gas formalism to the case with non-equal ranks of the gauge group, $M>0$.
As a result we have found non-trivial evidences that the partition functions for $m_1=m_2$ satisfies an infinite set of exact bilinear relations \eqref{conjecturedmABJMbilinearidentity}.
For $m_1=m_2=-\pi i(\nu-2a)/\nu$ with $\nu,a\in\mathbb{N}$, $1\le a\le \nu-1$ our proposal \eqref{conjecturedmABJMbilinearidentity} coincides with the higher rank generalization of the result in \cite{Bonelli:2017gdk} which relates the partition function of the ABJM theory without mass deformation to the $\tau$-function of the $q$-Painlev\'e $\text{III}_3$ (or affine $\text{SU}(2)$ $q$-Toda) equation.
Remarkably, we have observed that the bilinear relation \eqref{conjecturedmABJMbilinearidentity} is satisfied not only for $m_1=m_2=-\pi i(\nu-2a)/\nu$ but also for general $m_1=m_2\in \mathbb{C}$.
Hence \eqref{conjecturedmABJMbilinearidentity} also provides relations between the derivatives of the partition functions with respect to the mass parameter which gives some of the OPE coefficients for the stress tensor multiplet \cite{Agmon:2017xes}.

There are several possible future directions related to this work.
First, although in this paper we have found the relation only for $m_1=m_2$, it would be nice if we can generalize the result to $m_1\neq m_2$.
Second, we can consider 3d superconformal Chern-Simons theories with more general quivers.
It is known that the $S^3$ partition function of a quiver superconformal Chern-Simons theory with the quiver given by the affine Dynkin diagram of $A_n,D_n,E_n$ scales in the large $N$ limit as $-\log Z\sim N^{3/2}$ in the limit of $N\rightarrow\infty$ if the ranks of the gauge groups are proportional to the comarks of each node and the total Chern-Simons level weighted by the comarks vanishes \cite{Gulotta:2011vp}.
For the affine $A_n$ quivers and the affine $D_n$ quivers the fermi gas formalism was already found \cite{Marino:2011eh,Assel:2015hsa,Moriyama:2015jsa}, and for the affine $A_n$ quivers it was further extended to the case with mass deformations \cite{Nosaka:2015iiw} as well as to the cases with non-equal ranks \cite{Moriyama:2017gye,Kubo:2018cqw,Kubo:2019ejc,Kubo:2020qed} separately.
We expect that for the latter cases it is also possible to turn on both of the mass deformation and the rank deformation at the same time, and if so, it would be interesting to identify the corresponding $q$-difference equations and check them as the identities among the partition functions for different ranks.
It would also be interesting to consider the generalization which are natural from the viewpoint of the $q$-difference equations, such as the $q$-deformed Painlev\'e of the other types or the generalization of the affine $\text{SU}(\nu)$ Toda system to the Toda systems for other Lie groups.


\section*{Acknowlegement}
The author would like to thank Giulio Bonelli, Fran Globlek, Alba Grassi, Kazunobu Maruyoshi, Sanefumi Moriyama, Yuji Sugimoto and Alessandro Tanzini for valuable discussions and useful comments.
A part of the results in appendix \ref{app_listofexactvalues} was computed on the high performance computing facilities provided in SISSA (Ulysses) and in YITP (Sushiki server).

\appendix

\section{Proof of Cauchy-Vandermonde determinant formula \eqref{CauchyVdM}}
\label{sec_proofCVdM}

In this section we display a proof for the Cauchy-Vandermonde determinant formula \eqref{CauchyVdM} by using the Cauchy determinant formula for the equal rank
\begin{align}
\frac{\prod_{m<n}^{N+M}(a_m-a_n)\prod_{m<n}^{N+M}(b_m-b_n)}{\prod_{m=1}^{N+M}\prod_{n=1}^{N+M}(a_m+b_n)}=\det\Bigl[\frac{1}{a_m+b_n}\Bigr]_{m,n:(N+M)\times (N+M)},
\label{Cauchy}
\end{align}
as explained in \cite{Matsumoto:2013nya}.

First we take the limti $a_{N+M}\rightarrow\infty$ in the Cauchy determinant formula for rank $N+M$ \eqref{Cauchy}.
Under this limit the left-hand side becomes
\begin{align}
\frac{\prod_{m<n}^{N+M}(a_m-a_n)\prod_{m<n}^{N+M}(b_m-b_n)}{\prod_{m=1}^{N+M}\prod_{n=1}^{N+M}(a_m+b_n)}
&=\frac{\prod_{m<n}^{N+M-1}(a_m-a_n)\prod_{m<n}^{N+M}(b_m-b_n)}{\prod_{m=1}^{N+M-1}\prod_{n=1}^{N+M}(a_m+b_n)}\frac{\prod_{m=1}^{N+M-1}(a_m-a_{N+M})}{\prod_{n=1}^{N+M}(a_{N+M}-b_n)}\nonumber \\
&\mathop{\longrightarrow}^{a_{N+M}\rightarrow\infty} \frac{\prod_{m<n}^{N+M-1}(a_m-a_n)\prod_{m<n}^{N+M}(b_m-b_n)}{\prod_{m=1}^{N+M-1}\prod_{n=1}^{N+M}(a_m+b_n)}\frac{(-1)^{N+M-1}}{a_{N+M}}+{\cal O}(a_{N+M}^{-2}).
\label{lhs1}
\end{align}
To take the same limit of the right-hand side of \eqref{Cauchy} we expand the $(N+M)$-th row of the matrix as
\begin{align}
\det\Bigl[\frac{1}{a_m+b_n}\Bigr]_{m,n:(N+M)\times (N+M)}
&=\det\begin{pmatrix}
\Bigl[\frac{1}{a_m+b_n}\Bigr]_{m,n:(N+M-1)\times (N+M)}\vspace{0.2cm} \\
\Bigl[\frac{1}{a_{N+M}}\Bigl(1-\frac{b_n}{a_{N+M}}+\Bigl(\frac{b_n}{a_{N+M}}\Bigr)^2+\cdots\Bigr)\Bigr]_{n:N+M}
\end{pmatrix}\nonumber \\
&\mathop{\longrightarrow}^{a_{N+M}\rightarrow\infty}
\frac{1}{a_{N+M}}
\det\begin{pmatrix}
\Bigl[\frac{1}{a_m+b_n}\Bigr]_{m,n:(N+M-1)\times (N+M)}\vspace{0.2cm} \\
[1]_{N+M}
\end{pmatrix}+{\cal O}(a_{N+M}^{-2}).
\label{rhs1}
\end{align}
By equating the leading part of \eqref{lhs1} and \eqref{rhs1} we hence obtain
\begin{align}
\frac{\prod_{m<n}^{N+M-1}(a_m-a_n)\prod_{m<n}^{N+M}(b_m-b_n)}{\prod_{m=1}^{N+M-1}\prod_{n=1}^{N+M}(a_m+b_n)}=(-1)^{N+M-1}
\det\begin{pmatrix}
\Bigl[\frac{1}{a_m+b_n}\Bigr]_{m,n:(N+M-1)\times (N+M)}\vspace{0.2cm} \\
[1]_{N+M}
\end{pmatrix}.
\label{Cauchy2}
\end{align}
Now we further take the limit $a_{N+M-1}\rightarrow\infty$ in \eqref{Cauchy2}.
The calculation is the same as above, and from the left-hand side we obtain
\begin{align}
\frac{\prod_{m<n}^{N+M-1}(a_m-a_n)\prod_{m<n}^{N+M}(b_m-b_n)}{\prod_{m=1}^{N+M-1}\prod_{n=1}^{N+M}(a_m+b_n)}
&\mathop{\longrightarrow}^{a_{N+M-1}\rightarrow\infty} \frac{\prod_{m<n}^{N+M-2}(a_m-a_n)\prod_{m<n}^{N+M}(b_m-b_n)}{\prod_{m=1}^{N+M-2}\prod_{n=1}^{N+M}(a_m+b_n)}\frac{(-1)^{N+M-2}}{a_{N+M-1}^2}\nonumber \\
&\quad\quad\quad\quad\quad\quad\quad\quad +{\cal O}(a_{N+M-1}^{-3}).
\end{align}
From the right-hand side we obtain
\begin{align}
\det\begin{pmatrix}
\Bigl[\frac{1}{a_m+b_n}\Bigr]_{m,n:(N+M-1)\times (N+M)}\vspace{0.2cm}\\
[1]_{N+M}
\end{pmatrix}
&=
\det\begin{pmatrix}
\Bigl[\frac{1}{a_m+b_n}\Bigr]_{m,n:(N+M-2)\times (N+M)}\vspace{0.2cm} \\
\Bigl[\frac{1}{a_{N+M-1}}\Bigl(1-\frac{b_n}{a_{N+M-1}}+\Bigl(\frac{b_n}{a_{N+M-1}}\Bigr)^2+\cdots\Bigr)\Bigr]_{n:N+M}\vspace{0.2cm}\\
[1]_{N+M}
\end{pmatrix}\nonumber \\
&\mathop{\longrightarrow}^{a_{N+M-1}\rightarrow\infty}
-\frac{1}{a_{N+M-1}^2}
\det\begin{pmatrix}
\Bigl[\frac{1}{a_m+b_n}\Bigr]_{m,n:(N+M-2)\times (N+M)}\vspace{0.2cm} \\
[b_n]_{n:N+M}\vspace{0.2cm}\\
[1]_{N+M}
\end{pmatrix}\nonumber \\
&\quad\quad\quad\quad\quad\quad\quad\quad +{\cal O}(a_{N+M-1}^{-3}),
\end{align}
where in the expansion of the $(N+M-1)$-th row we picked up the sub-leading terms $-\frac{b_n}{a_{N+M-1}^2}$ instead of the leading terms $\frac{1}{a_{N+M-1}}$ since the leading terms are proportional to the $N+M$-th row and can be subtracted by the elementary row operation.
Hence from the leading part of \eqref{Cauchy2} in the limit $a_{N+M-1}\rightarrow\infty$ we obtain
\begin{align}
\frac{\prod_{m<n}^{N+M-2}(a_m-a_n)\prod_{m<n}^{N+M}(b_m-b_n)}{\prod_{m=1}^{N+M-2}\prod_{n=1}^{N+M}(a_m+b_n)}=
\Bigl(\prod_{\ell=1}^2(-1)^{N+M-\ell}(-1)^{\ell-1}\Bigr)
\det\begin{pmatrix}
\Bigl[\frac{1}{a_m+b_n}\Bigr]_{m,n:(N+M-2)\times (N+M)}\vspace{0.2cm} \\
[b_n]_{n:N+M}\vspace{0.2cm}\\
[1]_{N+M}
\end{pmatrix}.
\end{align}
By repeating the same procedures for $a_{N+M-2},a_{N+M-3},\cdots,a_{N+1}$ we obtain
\begin{align}
\frac{\prod_{i<j}^{N}(a_i-a_j)\prod_{m<n}^{N+M}(b_m-b_n)}{\prod_{i=1}^{N}\prod_{n=1}^{N+M}(a_i+b_n)}=
(-1)^{MN}
\det\begin{pmatrix}
\Bigl[\frac{1}{a_i+b_n}\Bigr]_{i,n:N\times (N+M)}\vspace{0.2cm} \\
[b_n^{M-r}]_{r,n:M\times (N+M)}
\end{pmatrix}.
\label{CauchyVdMnothyperbolic}
\end{align}

If we substitute $a_i=e^{x_i},b_m=e^{y_m}$ in \eqref{CauchyVdMnothyperbolic} we finally obtain the first line of the formulas \eqref{CauchyVdM}.
The second line of \eqref{CauchyVdM} is obtained if we instead substitute $a_i=e^{-x_i},b_m=e^{-y_m}$ to \eqref{CauchyVdMnothyperbolic}.


\section{List of exact values of $\text{tr}{\widehat\rho}^n$ and $\langle\!\langle 2\pi i\ell_r|{\widehat\rho}^n|x\rangle$}
\label{app_listofexactvalues}
As we explained in section \ref{sec_openstringformalism} the grand partition functions $\Xi_{k,M}(N,m_1,m_2)$, or its expansion coefficients $Z_{k,M}(N,m_1,m_2)$, can be obtained from $\text{Tr}{\widehat\rho}_0^n$ ($n=1,2,\cdots,N$) and $\langle\!\langle 2\pi i\ell_r|{\widehat\rho}_0^n|x\rangle$ ($n=0,1,\cdots,N$, $r=1,2,\cdots,M$) through \eqref{Xi_openstringformalism} and \eqref{Openfinal}.
Below we display the exact values of these quantities.

\subsection{$n\le 2$ without fixing $m_1,m_2$}
\label{app_listofexactvalueswithoutfixing}
\subsubsection{$n=0$}
\begin{align}
\langle\!\langle 2\pi i\ell_r|x\rangle=\frac{1}{\sqrt{k}}e^{\frac{\ell_rx}{k}}
\end{align}
where $\ell_r=M+1/2-r$.

\subsubsection{$n=1$}
\begin{align}
\text{Tr}{\widehat\rho}_0&=\frac{1}{4k\cosh\frac{m_1}{2}\cosh\frac{m_2}{2}},\nonumber \\
\langle\!\langle 2\pi i\ell_r|{\widehat\rho}_0|x\rangle&=\frac{1}{2\pi k\sqrt{k}}\frac{1}{1-(-1)^{p+q}e^{2\pi ip(\frac{\ell_r}{k}-\frac{im_2}{2\pi})}}\biggl[-\sum_{n=1}^p\frac{(-1)^ne^{(\frac{\ell_r}{k}-\frac{im_2}{2\pi})2\pi i(n-\frac{1}{2})}}{\cosh[\frac{1}{2k}(x+m_1k-2\pi i(n-\frac{1}{2}))]}\nonumber \\
&\quad 
-k\sum_{n=1}^q\frac{(-1)^ne^{(\frac{\ell_r}{k}-\frac{im_2}{2\pi})(x+m_1k+2\pi ik(n-\frac{1}{2}))}}{\cosh[\frac{1}{2}(x+m_1k+2\pi ik(n-\frac{1}{2}))]}
\biggr],
\end{align}
where for $\langle\!\langle 2\pi i\ell_r|{\widehat\rho}_0|x\rangle$ we have assumed $k=p/q\in\mathbb{Q}$ with $p,q$ coprime.

\subsubsection{$n=2$}
\begin{align}
\text{Tr}{\widehat\rho}_0^2&=\frac{1}{4k^2\sinh m_2}\frac{1}{1-(-1)^p\cosh[m_2p]}\biggl[
-\frac{ik}{\sinh m_1}\sum_{n=1}^q\sum_\pm\Bigl(\pm\frac{\sin[\frac{m_2k}{2\pi}(m_1\pm 2\pi i(n-\frac{1}{2}))]}{\sinh[\frac{k}{2}(m_1\pm 2\pi i(n-\frac{1}{2}))]}\Bigr)\nonumber \\
&\quad -\sum_{n=1}^{p-1}\frac{(-1)^n\sinh[m_2n]}{\cosh[\frac{1}{2k}(m_1k+2\pi in)]\cosh[\frac{1}{2k}(m_1k-2\pi in)]}
-\frac{(-1)^p\sinh[m_2p]}{2(\cosh\frac{m_1}{2})^2}
\biggr],\label{Trrho02} \\
\langle\!\langle 2\pi i\ell_r|{\widehat\rho}_0^2|x\rangle
&=\frac{1}{4k^2\sqrt{k}}\frac{1}{1-(-1)^{p+q}e^{2\pi ip(\frac{\ell_r}{k}-\frac{im_2}{2\pi})}}\biggl[\nonumber \\
&\quad \sum_{n=1}^p
\frac{(-1)^ne^{(\frac{\ell_r}{k}-\frac{im_2}{2\pi})2\pi i(n-\frac{1}{2})}}{1-(-1)^pe^{m_2p}}\biggl(\nonumber \\
&\quad\quad \sum_{m=1}^p\frac{(-1)^me^{m_2(m-\frac{1}{2})}}{\cosh[\frac{1}{2k}(x+m_1k-2\pi i(m-\frac{1}{2}))]\cosh[\frac{1}{2k}(m_1k-2\pi i(n-m))]}\nonumber \\
&\quad\quad +\sum_{m=1}^q\frac{ike^{-\frac{im_2}{2\pi}(x+m_1k+2\pi ik(m-\frac{1}{2}))}}{\cosh[\frac{1}{2}(x+m_1k+2\pi ik(m-\frac{1}{2}))]\sinh[\frac{1}{2k}(x+2m_1k-2\pi i(n-\frac{1}{2}))]}\nonumber \\
&\quad\quad +\Bigl(\sum_{m\in I_n}+\frac{1}{2}\sum_{m\in\bar{I}_n}\Bigr)\frac{-k(-1)^ne^{-\frac{im_2}{2\pi}(-m_1k+2\pi i(n-\frac{1}{2})+2\pi ik(m-\frac{1}{2}))}}{\sinh[\frac{1}{2}(m_1k-2\pi ik(m-\frac{1}{2}))]\sinh[\frac{1}{2k}(x+2m_1k-2\pi i(n-\frac{1}{2}))]}\biggr)\nonumber \\
&\quad +\sum_{n=1}^q\frac{k(-1)^ne^{(\frac{\ell_r}{k}-\frac{im_2}{2\pi})(m_1k+2\pi ik(n-\frac{1}{2}))}}{1-(-1)^qe^{(\frac{\ell_r}{k}-\frac{im_2}{\pi})2\pi ip}}\biggl(\nonumber \\
&\quad\quad \sum_{m=1}^p\frac{ie^{(\frac{\ell_r}{k}-\frac{im_2}{\pi})2\pi i(m-\frac{1}{2})}}{\sinh[\frac{1}{2}(m_1k+2\pi ik(n-\frac{1}{2}))]\cosh[\frac{1}{2k}(x+m_1k-2\pi i(m-\frac{1}{2}))]}\nonumber \\
&\quad\quad +\Bigl(\sum_{m\in J_n}+\frac{1}{2}\sum_{m\in\bar{J}_n}\Bigr)\frac{(-1)^ne^{(\frac{\ell_r}{k}-\frac{im_2}{\pi})(-m_1k-2\pi ik(n-\frac{1}{2})+2\pi i(m-\frac{1}{2}))}}{\sinh[\frac{1}{2}(m_1k+2\pi ik(n-\frac{1}{2}))]\sinh[\frac{1}{2k}(x+2m_1k-2\pi i(m-\frac{1}{2}))]}\nonumber \\
&\quad\quad +\sum_{m=1}^q\frac{k(-1)^me^{(\frac{\ell_r}{k}-\frac{im_2}{\pi})(x+m_1k+2\pi ik(m-\frac{1}{2}))}}{\cosh[\frac{1}{2}(x+m_1k+2\pi ik(m-\frac{1}{2}))]\cosh[\frac{1}{2}(x+2m_1k+2\pi ik(n+m-1))]}
\biggr) \biggr]
\end{align}
Here again we have assumed $k=p/q\in\mathbb{Q}$ with $p,q$ coprime\footnote{
Note that $\text{Tr}{\widehat\rho}_0^2$ were already obtained in \cite{Anderson:2015ioa,Russo:2015exa} for $k\in\mathbb{N}$.  
We have checked that the expression \eqref{Trrho02} is consistent with the known result.
}, and the index sets $I_n,\bar{I}_n,J_n,\bar{J}_n$ are defined as
\begin{align}
I_n&=\Bigl\{m\in\mathbb{Z}\,|\,0<2\pi\Bigl(n-\frac{1}{2}\Bigr)+2\pi k\Bigl(m-\frac{1}{2}\Bigr)<2\pi p\Bigr\},\nonumber \\
\bar{I}_n&=\Bigl\{m\in\mathbb{Z}\,|\,2\pi\Bigl(n-\frac{1}{2}\Bigr)+2\pi k\Bigl(m-\frac{1}{2}\Bigr)=0,2\pi p\Bigr\},\nonumber \\
J_n&=\Bigl\{m\in\mathbb{Z}\,|\,0<-2\pi k\Bigl(n-\frac{1}{2}\Bigr)+2\pi\Bigl(m-\frac{1}{2}\Bigr)<2\pi p\Bigr\},\nonumber \\
\bar{J}_n&=\Bigl\{m\in\mathbb{Z}\,|\,-2\pi k\Bigl(n-\frac{1}{2}\Bigr)+2\pi\Bigl(m-\frac{1}{2}\Bigr)=0,2\pi p\Bigr\}.
\end{align}

\subsection{$n\ge 3$ with fixing $m_1,m_2$}
Below we display $Z_{k,M}(N,m_1,m_2)$ instead of $\langle\!\langle 2\pi i\ell_r|{\widehat\rho}_0^n|x\rangle$ whose expressions are unnecessarily complicated.
\subsubsection{$Z_{k,M}(N)$ for $m_1=m_2=-\frac{\pi i}{3}$ ($\nu=3, a=1$), $k=6$, $M\le 3$}
\begin{align}
Z_{6,0}(3)&=\frac{648+144\sqrt{3}\pi-145\pi^2}{279936\pi^2},\nonumber \\
Z_{6,1}(3)&=\frac{(1+i)(-648\sqrt{3}+(10368-21060\sqrt{3}i)\pi-(1854\sqrt{3}+63072i)\pi^2+(8+13725\sqrt{3}i)\pi^3)}{40310784\pi^3},\nonumber \\
Z_{6,2}(3)&=\frac{
-33048-9720\sqrt{3}i+(142236\sqrt{3}-210924i)\pi+(192510-26730\sqrt{3}i)\pi^2+(-49175\sqrt{3}+36651i)\pi^3
}
{241864704\pi^3},\nonumber \\
Z_{6,3}(3)&=-\frac{(1+i)(-4968+1656\sqrt{3}i+913(\sqrt{3}-i)\pi)}{6718464\pi}.
\end{align}

\subsubsection{$Z_{k,M}(N)$ for $m_1=m_2=-\frac{\pi i}{2}$ ($\nu=4, a=1$), $k=2$, $M\le 2$}
%
\begin{align}
Z_{2,0}(3)&=\frac{\sqrt{2}+(1-\sqrt{2})\pi}{64\pi},\nonumber \\
Z_{2,1}(3)&=\frac{-2-2i-2\sqrt{2}+(3-5i+(-1+4i)\sqrt{2})\pi}{256\pi},\nonumber \\
Z_{2,2}(3)&=\frac{(1+i)(2+(-2+\sqrt{2})\pi)}{128\pi},\nonumber \\
Z_{2,0}(4)&=\frac{-12+(1+2\sqrt{2})\pi}{1024\pi},\nonumber \\
Z_{2,1}(4)&=-\frac{(1+i)(4+4\pi+(-13+8\sqrt{2})\pi^2)}{4096\pi^2},\nonumber \\
Z_{2,2}(4)&=\frac{i(-12+(1+2\sqrt{2})\pi)}{1024\pi},\nonumber \\
Z_{2,0}(5)&=\frac{1+(5-\sqrt{2})\pi+(3-3\sqrt{2})\pi^2}{1024\pi^2},\nonumber \\
Z_{2,1}(5)&=\frac{-4+4i+4\sqrt{2}+(68+60i-(4-64i)\sqrt{2})\pi+(21-37i-(29+8i)\sqrt{2})\pi^2}{16384\pi^2},\nonumber \\
Z_{2,2}(5)&=-\frac{(1-i)(\sqrt{2}+(-2+5\sqrt{2})\pi+(-6+3\sqrt{2})\pi^2)}{2048\pi^2},\nonumber \\
Z_{2,0}(6)&=\frac{8\sqrt{2}+(72+28\sqrt{2})\pi+(10-33\sqrt{2})\pi^2}{32768\pi^2},\nonumber \\
Z_{2,1}(6)&=-\frac{i(-24\sqrt{2}-108\sqrt{2}\pi+(288+1666\sqrt{2})\pi^2+(-1296+333\sqrt{2})\pi^3)}{1179648\pi^3},\nonumber \\
Z_{2,2}(6)&=\frac{-8\sqrt{2}-(72+28\sqrt{2})\pi-10\pi^2+33\sqrt{2}\pi^2}{32768\pi^2},\nonumber \\
Z_{2,0}(7)&=\frac{-6\sqrt{2}+(18+72\sqrt{2})\pi-(270+287\sqrt{2})\pi^2+(216-9\sqrt{2})\pi^3}{294912\pi^3},\nonumber \\
Z_{2,1}(7)&=
\frac{1}{4718592\pi^3}
\Bigl(
(24 - 24 i) - 24 \sqrt{2}i
+( (900 + 1116 i) - (1008 - 108 i) \sqrt{2})\pi\nonumber \\
&\quad +( - (1152 - 3682 i)  - (2530 - 4834 i) \sqrt{2})\pi^2
+( - (711 + 2241 i) + (1476 - 765 i) \sqrt{2})\pi^3\Bigr),\nonumber \\
Z_{2,2}(7)&=
-\frac{(1+i)
(-6
 + (72 + 9 \sqrt{2})\pi
+( - 287 - 135 \sqrt{2})\pi^2
+(- 9 \pi^3 + 108 \sqrt{2})\pi^3)
}
{294912\pi^3}.
\end{align}



\subsubsection{$Z_{k,M}(N)$ for $m_1=m_2=-\frac{\pi i}{2}$ ($\nu=4, a=1$), $k=4$, $M\le 2$}
\begin{align}
Z_{4,0}(3)&=\frac{2-10\pi+3\pi^2}{1024\pi^2},\nonumber \\
Z_{4,1}(3)&=\frac{6-6i+(36+54i)\pi-(40-238i)\pi^2+(9-81i)\pi^3}{36864\sqrt{2}\pi^3},\nonumber \\
Z_{4,2}(3)&=\frac{i(3+28\pi^2-9\pi^3)}{9216\pi^3},\nonumber \\
Z_{4,0}(4)&=\frac{70+6\pi-9\pi^2}{49152\pi^2},\nonumber \\
Z_{4,1}(4)&=\frac{-3+3i-(60+48i)\pi+(1040-626i)\pi^2+(382-418i)\pi^3-(225-198i)\pi^4}{589824\sqrt{2}\pi^4},\nonumber \\
Z_{4,2}(4)&=-\frac{(1-i)(48-708\pi-656\pi^2+279\pi^3)}{1179648\pi^3}.
\end{align}


\begin{thebibliography}{99}
\bibitem{Klebanov:1996un}
  I.~R.~Klebanov and A.~A.~Tseytlin,
  ``Entropy of near extremal black p-branes,''
  Nucl. Phys. B \textbf{475} (1996), 164-178
  doi:10.1016/0550-3213(96)00295-7
  [arXiv:hep-th/9604089 [hep-th]].
%
\bibitem{Schwarz:2004yj}
  J.~H.~Schwarz,
  ``Superconformal Chern-Simons theories,''
  JHEP \textbf{11} (2004), 078
  doi:10.1088/1126-6708/2004/11/078
  [arXiv:hep-th/0411077 [hep-th]].
%
\bibitem{Hosomichi:2008jd}
  K.~Hosomichi, K.~M.~Lee, S.~Lee, S.~Lee and J.~Park,
  ``N=4 Superconformal Chern-Simons Theories with Hyper and Twisted Hyper Multiplets,''
  JHEP \textbf{07} (2008), 091
  doi:10.1088/1126-6708/2008/07/091
  [arXiv:0805.3662 [hep-th]].
%
\bibitem{Aharony:2008ug}
  O.~Aharony, O.~Bergman, D.~L.~Jafferis and J.~Maldacena,
  ``N=6 superconformal Chern-Simons-matter theories, M2-branes and their gravity duals,''
  JHEP \textbf{10} (2008), 091
  doi:10.1088/1126-6708/2008/10/091
  [arXiv:0806.1218 [hep-th]].
%
\bibitem{Herzog:2010hf}
C.~P.~Herzog, I.~R.~Klebanov, S.~S.~Pufu and T.~Tesileanu,
``Multi-Matrix Models and Tri-Sasaki Einstein Spaces,''
Phys. Rev. D \textbf{83} (2011), 046001
doi:10.1103/PhysRevD.83.046001
[arXiv:1011.5487 [hep-th]].
%
\bibitem{Kapustin:2009kz}
  A.~Kapustin, B.~Willett and I.~Yaakov,
  ``Exact Results for Wilson Loops in Superconformal Chern-Simons Theories with Matter,''
  JHEP \textbf{03} (2010), 089
  doi:10.1007/JHEP03(2010)089
  [arXiv:0909.4559 [hep-th]].
%
\bibitem{Drukker:2010nc}
N.~Drukker, M.~Marino and P.~Putrov,
``From weak to strong coupling in ABJM theory,''
Commun. Math. Phys. \textbf{306} (2011), 511-563
doi:10.1007/s00220-011-1253-6
[arXiv:1007.3837 [hep-th]].
%
\bibitem{Suyama:2009pd}
T.~Suyama,
``On Large N Solution of ABJM Theory,''
Nucl. Phys. B \textbf{834} (2010), 50-76
doi:10.1016/j.nuclphysb.2010.03.011
[arXiv:0912.1084 [hep-th]].
%
\bibitem{Drukker:2009hy}
  N.~Drukker and D.~Trancanelli,
  ``A Supermatrix model for N=6 super Chern-Simons-matter theory,''
  JHEP \textbf{02} (2010), 058
  doi:10.1007/JHEP02(2010)058
  [arXiv:0912.3006 [hep-th]].
%
\bibitem{Marino:2009jd}
M.~Marino and P.~Putrov,
``Exact Results in ABJM Theory from Topological Strings,''
JHEP \textbf{06} (2010), 011
doi:10.1007/JHEP06(2010)011
[arXiv:0912.3074 [hep-th]].
%
\bibitem{Drukker:2011zy}
N.~Drukker, M.~Marino and P.~Putrov,
``Nonperturbative aspects of ABJM theory,''
JHEP \textbf{11} (2011), 141
doi:10.1007/JHEP11(2011)141
[arXiv:1103.4844 [hep-th]].
%
\bibitem{Suyama:2011yz}
T.~Suyama,
``Eigenvalue Distributions in Matrix Models for Chern-Simons-matter Theories,''
Nucl. Phys. B \textbf{856} (2012), 497-527
doi:10.1016/j.nuclphysb.2011.11.013
[arXiv:1106.3147 [hep-th]].
%
\bibitem{Fuji:2011km}
H.~Fuji, S.~Hirano and S.~Moriyama,
``Summing Up All Genus Free Energy of ABJM Matrix Model,''
JHEP \textbf{08} (2011), 001
doi:10.1007/JHEP08(2011)001
[arXiv:1106.4631 [hep-th]].
%
\bibitem{Okuyama:2011su}
  K.~Okuyama,
  ``A Note on the Partition Function of ABJM theory on $S^3$,''
  Prog. Theor. Phys. \textbf{127} (2012), 229-242
  doi:10.1143/PTP.127.229
  [arXiv:1110.3555 [hep-th]].
%
\bibitem{Marino:2011eh}
  M.~Marino and P.~Putrov,
  ``ABJM theory as a Fermi gas,''
  J.\ Stat.\ Mech.\  {\bf 1203} (2012) P03001
  doi:10.1088/1742-5468/2012/03/P03001
  [arXiv:1110.4066 [hep-th]].
%
\bibitem{Hatsuda:2012hm}
Y.~Hatsuda, S.~Moriyama and K.~Okuyama,
``Exact Results on the ABJM Fermi Gas,''
JHEP \textbf{10} (2012), 020
doi:10.1007/JHEP10(2012)020
[arXiv:1207.4283 [hep-th]].
%
\bibitem{Putrov:2012zi}
  P.~Putrov and M.~Yamazaki,
  ``Exact ABJM Partition Function from TBA,''
  Mod.\ Phys.\ Lett.\ A {\bf 27} (2012) 1250200
  doi:10.1142/S0217732312502008
  [arXiv:1207.5066 [hep-th]].
%
\bibitem{Hatsuda:2012dt}
  Y.~Hatsuda, S.~Moriyama and K.~Okuyama,
  ``Instanton Effects in ABJM Theory from Fermi Gas Approach,''
  JHEP \textbf{01} (2013), 158
  doi:10.1007/JHEP01(2013)158
  [arXiv:1211.1251 [hep-th]].
%
\bibitem{Calvo:2012du}
F.~Calvo and M.~Marino,
``Membrane instantons from a semiclassical TBA,''
JHEP \textbf{05} (2013), 006
doi:10.1007/JHEP05(2013)006
[arXiv:1212.5118 [hep-th]].
%
\bibitem{Hatsuda:2013gj}
Y.~Hatsuda, S.~Moriyama and K.~Okuyama,
``Instanton Bound States in ABJM Theory,''
JHEP \textbf{05} (2013), 054
doi:10.1007/JHEP05(2013)054
[arXiv:1301.5184 [hep-th]].
%
\bibitem{Hatsuda:2013oxa}
Y.~Hatsuda, M.~Marino, S.~Moriyama and K.~Okuyama,
``Non-perturbative effects and the refined topological string,''
JHEP \textbf{09} (2014), 168
doi:10.1007/JHEP09(2014)168
[arXiv:1306.1734 [hep-th]].
%
\bibitem{Honda:2014npa}
  M.~Honda and K.~Okuyama,
  ``Exact results on ABJ theory and the refined topological string,''
  JHEP \textbf{08} (2014), 148
  doi:10.1007/JHEP08(2014)148
  [arXiv:1405.3653 [hep-th]].
%
\bibitem{Bonelli:2016idi}
  G.~Bonelli, A.~Grassi and A.~Tanzini,
  ``Seiberg–Witten theory as a Fermi gas,''
  Lett. Math. Phys. \textbf{107} (2017) no.1, 1-30
  doi:10.1007/s11005-016-0893-z
  [arXiv:1603.01174 [hep-th]].
%
\bibitem{Bonelli:2017ptp}
  G.~Bonelli, A.~Grassi and A.~Tanzini,
  ``New results in $\mathcal{N}=2$ theories from non-perturbative string,''
  Annales Henri Poincare \textbf{19} (2018) no.3, 743-774
  doi:10.1007/s00023-017-0643-5
  [arXiv:1704.01517 [hep-th]].
%
\bibitem{Bonelli:2017gdk}
  G.~Bonelli, A.~Grassi and A.~Tanzini,
  ``Quantum curves and $q$-deformed Painlevé equations,''
  Lett.\ Math.\ Phys.\  {\bf 109} (2019) no.9,  1961
  doi:10.1007/s11005-019-01174-y
  [arXiv:1710.11603 [hep-th]].
%
\bibitem{Zamolodchikov:1994uw}
A.~B.~Zamolodchikov,
``Painleve III and 2-d polymers,''
Nucl. Phys. B \textbf{432} (1994), 427-456
doi:10.1016/0550-3213(94)90029-9
[arXiv:hep-th/9409108 [hep-th]].
%
\bibitem{Grassi:2014uua}
A.~Grassi, Y.~Hatsuda and M.~Marino,
``Quantization conditions and functional equations in ABJ(M) theories,''
J. Phys. A \textbf{49} (2016) no.11, 115401
doi:10.1088/1751-8113/49/11/115401
[arXiv:1410.7658 [hep-th]].
%
\bibitem{Hosomichi:2008jb}
K.~Hosomichi, K.~M.~Lee, S.~Lee, S.~Lee and J.~Park,
``N=5,6 Superconformal Chern-Simons Theories and M2-branes on Orbifolds,''
JHEP \textbf{09} (2008), 002
doi:10.1088/1126-6708/2008/09/002
[arXiv:0806.4977 [hep-th]].
%
\bibitem{Gomis:2008vc}
J.~Gomis, D.~Rodriguez-Gomez, M.~Van Raamsdonk and H.~Verlinde,
``A Massive Study of M2-brane Proposals,''
JHEP \textbf{09} (2008), 113
doi:10.1088/1126-6708/2008/09/113
[arXiv:0807.1074 [hep-th]].
%
\bibitem{Jafferis:2011zi}
  D.~L.~Jafferis, I.~R.~Klebanov, S.~S.~Pufu and B.~R.~Safdi,
  ``Towards the F-Theorem: N=2 Field Theories on the Three-Sphere,''
  JHEP \textbf{06} (2011), 102
  doi:10.1007/JHEP06(2011)102
  [arXiv:1103.1181 [hep-th]].
%
\bibitem{Anderson:2015ioa}
  L.~Anderson and J.~G.~Russo,
  ``ABJM Theory with mass and FI deformations and Quantum Phase Transitions,''
  JHEP \textbf{05} (2015), 064
  doi:10.1007/JHEP05(2015)064
  [arXiv:1502.06828 [hep-th]].
%
\bibitem{Nosaka:2015bhf}
T.~Nosaka, K.~Shimizu and S.~Terashima,
``Large N behavior of mass deformed ABJM theory,''
JHEP \textbf{03} (2016), 063
doi:10.1007/JHEP03(2016)063
[arXiv:1512.00249 [hep-th]].
%
\bibitem{Nosaka:2016vqf}
T.~Nosaka, K.~Shimizu and S.~Terashima,
``Mass Deformed ABJM Theory on Three Sphere in Large N limit,''
JHEP \textbf{03} (2017), 121
doi:10.1007/JHEP03(2017)121
[arXiv:1608.02654 [hep-th]].
%
\bibitem{Nosaka:2015iiw}
  T.~Nosaka,
  ``Instanton effects in ABJM theory with general R-charge assignments,''
  JHEP {\bf 1603} (2016) 059
  doi:10.1007/JHEP03(2016)059
  [arXiv:1512.02862 [hep-th]].
%
\bibitem{Honda:2018pqa}
  M.~Honda, T.~Nosaka, K.~Shimizu and S.~Terashima,
  ``Supersymmetry Breaking in a Large $N$ Gauge Theory with Gravity Dual,''
  JHEP {\bf 1903} (2019) 159
  doi:10.1007/JHEP03(2019)159
  [arXiv:1807.08874 [hep-th]].
%
\bibitem{Poland:2010wg}
D.~Poland and D.~Simmons-Duffin,
``Bounds on 4D Conformal and Superconformal Field Theories,''
JHEP \textbf{05} (2011), 017
doi:10.1007/JHEP05(2011)017
[arXiv:1009.2087 [hep-th]].
%
\bibitem{ElShowk:2012hu}
S.~El-Showk and M.~F.~Paulos,
``Bootstrapping Conformal Field Theories with the Extremal Functional Method,''
Phys. Rev. Lett. \textbf{111} (2013) no.24, 241601
doi:10.1103/PhysRevLett.111.241601
[arXiv:1211.2810 [hep-th]].
%
\bibitem{El-Showk:2014dwa}
S.~El-Showk, M.~F.~Paulos, D.~Poland, S.~Rychkov, D.~Simmons-Duffin and A.~Vichi,
``Solving the 3d Ising Model with the Conformal Bootstrap II. c-Minimization and Precise Critical Exponents,''
J. Stat. Phys. \textbf{157} (2014), 869
doi:10.1007/s10955-014-1042-7
[arXiv:1403.4545 [hep-th]].
%
\bibitem{Chester:2014fya}
S.~M.~Chester, J.~Lee, S.~S.~Pufu and R.~Yacoby,
``The $ \mathcal{N}=8 $ superconformal bootstrap in three dimensions,''
JHEP \textbf{09} (2014), 143
doi:10.1007/JHEP09(2014)143
[arXiv:1406.4814 [hep-th]].
%
\bibitem{Agmon:2017xes}
  N.~B.~Agmon, S.~M.~Chester and S.~S.~Pufu,
  ``Solving M-theory with the Conformal Bootstrap,''
  JHEP \textbf{06} (2018), 159
  doi:10.1007/JHEP06(2018)159
  [arXiv:1711.07343 [hep-th]].
%
\bibitem{Chester:2018aca}
S.~M.~Chester, S.~S.~Pufu and X.~Yin,
``The M-Theory S-Matrix From ABJM: Beyond 11D Supergravity,''
JHEP \textbf{08} (2018), 115
doi:10.1007/JHEP08(2018)115
[arXiv:1804.00949 [hep-th]].
%
\bibitem{Binder:2018yvd}
D.~J.~Binder, S.~M.~Chester and S.~S.~Pufu,
``Absence of $D^4 R^4$ in M-Theory From ABJM,''
JHEP \textbf{04} (2020), 052
doi:10.1007/JHEP04(2020)052
[arXiv:1808.10554 [hep-th]].
%
\bibitem{Binder:2019mpb}
D.~J.~Binder, S.~M.~Chester and S.~S.~Pufu,
``AdS$_{4}$/CFT$_{3}$ from weak to strong string coupling,''
JHEP \textbf{01} (2020), 034
doi:10.1007/JHEP01(2020)034
[arXiv:1906.07195 [hep-th]].
%
\bibitem{Chester:2020jay}
  S.~M.~Chester, R.~R.~Kalloor and A.~Sharon,
  ``3d $\mathcal{N}=4$ OPE Coefficients from Fermi Gas,''
  [arXiv:2004.13603 [hep-th]].
%
\bibitem{Binder:2020ckj}
D.~J.~Binder, S.~M.~Chester, M.~Jerdee and S.~S.~Pufu,
``The 3d $\mathcal{N}=6$ Bootstrap: From Higher Spins to Strings to Membranes,''
[arXiv:2011.05728 [hep-th]].
%
\bibitem{Matsumoto:2013nya}
  S.~Matsumoto and S.~Moriyama,
  ``ABJ Fractional Brane from ABJM Wilson Loop,''
  JHEP {\bf 1403} (2014) 079
  doi:10.1007/JHEP03(2014)079
  [arXiv:1310.8051 [hep-th]].
%
\bibitem{Bershtein:2018srt}
  M.~Bershtein, P.~Gavrylenko and A.~Marshakov,
  ``Cluster Toda chains and Nekrasov functions,''
  Theor. Math. Phys. \textbf{198} (2019) no.2, 157-188
  doi:10.1134/S0040577919020016
  [arXiv:1804.10145 [math-ph]].
%
\bibitem{Aharony:2008gk}
O.~Aharony, O.~Bergman, D.~L.~Jafferis,
``Fractional M2-branes,''
JHEP \textbf{11} (2008), 043
doi:10.1088/1126-6708/2008/11/043
[arXiv:0807.4924 [hep-th]].
%
\bibitem{Freedman:2013oja}
D.~Z.~Freedman and S.~S.~Pufu,
``The holography of $F$-maximization,''
JHEP \textbf{03} (2014), 135
doi:10.1007/JHEP03(2014)135
[arXiv:1302.7310 [hep-th]].
%
\bibitem{Awata:2012jb}
H.~Awata, S.~Hirano and M.~Shigemori,
``The Partition Function of ABJ Theory,''
PTEP \textbf{2013} (2013), 053B04
doi:10.1093/ptep/ptt014
[arXiv:1212.2966 [hep-th]].
%
\bibitem{Honda:2013pea}
M.~Honda,
``Direct derivation of ''mirror'' ABJ partition function,''
JHEP \textbf{12} (2013), 046
doi:10.1007/JHEP12(2013)046
[arXiv:1310.3126 [hep-th]].
%
\bibitem{Bershtein:2018zcz}
M.~Bershtein and A.~Shchechkin,
``Painlev\'e equations from Nakajima\textendash{}Yoshioka blowup relations,''
Lett. Math. Phys. \textbf{109} (2019) no.11, 2359-2402
doi:10.1007/s11005-019-01198-4
[arXiv:1811.04050 [math-ph]].
%
\bibitem{Nakajima:2005fg}
  H.~Nakajima and K.~Yoshioka,
  ``Instanton counting on blowup. II. K-theoretic partition function,''
  [arXiv:math/0505553 [math.AG]].
%
\bibitem{Kashaev:2015wia}
  R.~Kashaev, M.~Marino and S.~Zakany,
  ``Matrix models from operators and topological strings, 2,''
  Annales Henri Poincare \textbf{17} (2016) no.10, 2741-2781
  doi:10.1007/s00023-016-0471-z
  [arXiv:1505.02243 [hep-th]].
%
\bibitem{Tracy:1995ax}
  C.~A.~Tracy and H.~Widom,
  ``Proofs of two conjectures related to the thermodynamic Bethe ansatz,''
  Commun. Math. Phys. \textbf{179} (1996), 667-680
  doi:10.1007/BF02100102
  [arXiv:solv-int/9509003 [nlin.SI]].
%
\bibitem{Gulotta:2011vp}
D.~R.~Gulotta, J.~P.~Ang and C.~P.~Herzog,
``Matrix Models for Supersymmetric Chern-Simons Theories with an ADE Classification,''
JHEP \textbf{01} (2012), 132
doi:10.1007/JHEP01(2012)132
[arXiv:1111.1744 [hep-th]].
%
\bibitem{Russo:2015exa}
  J.~G.~Russo and G.~A.~Silva,
  ``Exact partition function in $U(2)\times U(2)$ ABJM theory deformed by mass and Fayet-Iliopoulos terms,''
  JHEP \textbf{12} (2015), 092
  doi:10.1007/JHEP12(2015)092
  [arXiv:1510.02957 [hep-th]].
%
\bibitem{Assel:2015hsa}
B.~Assel, N.~Drukker and J.~Felix,
``Partition functions of 3d $\hat D$-quivers and their mirror duals from 1d free fermions,''
JHEP \textbf{08} (2015), 071
doi:10.1007/JHEP08(2015)071
[arXiv:1504.07636 [hep-th]].
%
\bibitem{Moriyama:2015jsa}
S.~Moriyama and T.~Nosaka,
``Superconformal Chern-Simons Partition Functions of Affine D-type Quiver from Fermi Gas,''
JHEP \textbf{09} (2015), 054
doi:10.1007/JHEP09(2015)054
[arXiv:1504.07710 [hep-th]].
%
\bibitem{Moriyama:2017gye}
  S.~Moriyama, S.~Nakayama and T.~Nosaka,
  ``Instanton Effects in Rank Deformed Superconformal Chern-Simons Theories from Topological Strings,''
  JHEP \textbf{08} (2017), 003
  doi:10.1007/JHEP08(2017)003
  [arXiv:1704.04358 [hep-th]].
%
\bibitem{Kubo:2018cqw}
  N.~Kubo, S.~Moriyama and T.~Nosaka,
  ``Symmetry Breaking in Quantum Curves and Super Chern-Simons Matrix Models,''
  JHEP \textbf{01} (2019), 210
  doi:10.1007/JHEP01(2019)210
  [arXiv:1811.06048 [hep-th]].
%
\bibitem{Kubo:2019ejc}
  N.~Kubo and S.~Moriyama,
  ``Hanany-Witten Transition in Quantum Curves,''
  JHEP \textbf{12} (2019), 101
  doi:10.1007/JHEP12(2019)101
  [arXiv:1907.04971 [hep-th]].
%
\bibitem{Kubo:2020qed}
  N.~Kubo,
  ``Fermi gas approach to general rank theories and quantum curves,''
  [arXiv:2007.08602 [hep-th]].
%
\end{thebibliography}
\end{document}